\documentclass[twocolumn,pra,floatfix,superscriptaddress,nobibnotes]{revtex4-2} 
\pdfoutput=1 
\usepackage{amsmath}
\usepackage{amsfonts}
\usepackage{amssymb}
\usepackage{physics}
\usepackage{graphicx}
\usepackage{bbm}
\usepackage[dvipsnames]{xcolor}
\usepackage{braket}
\usepackage{mathrsfs}
\usepackage{lipsum} 
\usepackage{cancel}
\usepackage{xr}
\usepackage{soul}
\usepackage{array,
                booktabs,
                multirow}
\usepackage{mathtools}
\usepackage{hyperref}
\allowdisplaybreaks

\newcommand{\blue}{\textcolor{blue}}

\newcommand\approxRWA{\mathrel{\overset{\makebox[0pt]{\mbox{\normalfont\tiny\sffamily RWA }}}{\approx}}}

\makeatletter
\DeclareFontEncoding{LS1}{}{}
\DeclareFontSubstitution{LS1}{stix}{m}{n}
\DeclareMathAlphabet{\mathscr}{LS1}{stixscr}{m}{n}

\def \id{\mathbbm{\hat 1}}

\def \a{\hat a}
\def \c{\hat a^\dagger}

\def \I{\hat \sigma_0}
\def \X{\hat \sigma_x}
\def \Y{\hat \sigma_y}
\def \Z{\hat \sigma_z}
\def \p{\hat \sigma_{+}}
\def \m{\hat \sigma_{-}}

\def \HEsc{\hat H_{\footnotesize{\text{c/s},\text{exact}}}}
\def \HR{\hat H_{\text{red}}}

\def \Xj{\hat \sigma^{(j)}_x}
\def \Yj{\hat \sigma^{(j)}_y}
\def \Zj{\hat \sigma^{(j)}_z}
\def \pj{\hat \sigma_{+}^{(j)}}
\def \mj{\hat \sigma_{-}^{(j)}}

\def \SR{\hat S_{\text{red}}}
\def \SE{\hat S_{\footnotesize{\text{c/s},\text{exact}}}}

\def \stage{b}

\newcommand{\tiz}[1]{\tau_{\text{ini}}(#1)}
\newcommand{\tfz}[1]{\tau_{\text{fin}}(#1)}

\def \opvec{\hat \Theta}
\def \Hcoer{h_{\text{red}}}
\def \Hcoee{h_{\footnotesize{\text{c/s},\text{exact}}}}

\newcommand{\Ie}[1]{I^{(#1)}_{\footnotesize{\text{c/s},\text{exact}}}}
\newcommand{\Ir}[1]{I^{(#1)}_{\text{red}}}
\newcommand{\opIe}[1]{\hat I^{(#1)}_{\footnotesize{\text{c/s},\text{exact}}}}
\newcommand{\opIr}[1]{\hat I^{(#1)}_{\text{red}}}
\newcommand{\Ies}[1]{I^{(#1)}_{\footnotesize{\text{c/s},\text{exact}}}}
\newcommand{\Iec}[1]{I^{(#1)}_{\footnotesize{\text{c/s},\text{exact}}}}

\begin{document}

\title{High-fidelity regimes of resonator-mediated controlled-Z gates between quantum-dot qubits}

\author{Guangzhao \surname{Yang}}
\affiliation{School of Physical and Mathematical Sciences, Nanyang Technological University, 21 Nanyang Link, Singapore 637371, Singapore}
\author{Marek \surname{Gluza}}
\affiliation{School of Physical and Mathematical Sciences, Nanyang Technological University, 21 Nanyang Link, Singapore 637371, Singapore}
\author{Si Yan \surname{Koh}}
\affiliation{School of Physical and Mathematical Sciences, Nanyang Technological University, 21 Nanyang Link, Singapore 637371, Singapore}
\author{Kelvin \surname{Onggadinata}}
\affiliation{School of Physical and Mathematical Sciences, Nanyang Technological University, 21 Nanyang Link, Singapore 637371, Singapore}
\author{Calvin Pei Yu Wong}
\affiliation{Institute of Materials Research and Engineering (IMRE), Agency for Science, Technology and Research (A*STAR), Singapore 138634, Singapore}
\author{Kuan Eng Johnson Goh}
\affiliation{School of Physical and Mathematical Sciences, Nanyang Technological University, 21 Nanyang Link, Singapore 637371, Singapore}
\affiliation{Institute of Materials Research and Engineering (IMRE), Agency for Science, Technology and Research (A*STAR), Singapore 138634, Singapore}
\affiliation{Department of Physics, Faculty of Science, National University of Singapore, Singapore 117551, Singapore}
\author{Bent Weber}
\affiliation{School of Physical and Mathematical Sciences, Nanyang Technological University, 21 Nanyang Link, Singapore 637371, Singapore}
\author{Hui Khoon Ng}
\affiliation{Yale-NUS College, Singapore 138527, Singapore}
\affiliation{Centre for Quantum Technologies, National University of Singapore, Singapore 117543, Singapore}
\author{Teck Seng \surname{Koh}}
\thanks{Corresponding author:~\href{mailto:kohteckseng@ntu.edu.sg}{kohteckseng@ntu.edu.sg}}
\affiliation{School of Physical and Mathematical Sciences, Nanyang Technological University, 21 Nanyang Link, Singapore 637371, Singapore}
\email{kohteckseng@ntu.edu.sg}

\date{\today}

\begin{abstract}
Semiconductor double quantum dot (DQD) qubits coupled via superconducting microwave resonators provide a powerful means of long-range manipulation of the qubits' spin and charge degrees of freedom. Quantum gates can be implemented by parametrically driving the qubits while their transition frequencies are detuned from the resonator frequency. Long-range two-qubit controlled-Z (CZ) gates have been proposed for the DQD spin qubit within the rotating-wave approximation (RWA). Rapid gates demand strong coupling, but RWA breaks down when coupling strengths become significant relative to system frequencies. Therefore, understanding the errors arising from approximations used is critical for high-fidelity operation. Here, we go beyond RWA to study CZ gate fidelity for both DQD spin and charge qubits. We propose a novel parametric drive on the charge qubit that produces smaller errors and show that the fidelity of the CZ gate outperforms its spin counterpart, resulting in a much smaller fidelity loss of $0.05\%$ compared to $0.80\%$ for the spin qubit, and greater robustness against qubit dephasing and photon loss. We find that drive amplitude -- a parameter dropped in RWA -- is critical for optimizing fidelity, with the charge qubit exhibiting better tolerance to drive amplitude variations. Our results demonstrate the necessity of going beyond RWA in understanding how long-range gates can be realized in DQD qubits, with charge qubits offering considerable advantages in high-fidelity operation.
\end{abstract}

\maketitle

\section{Introduction}\label{sec:Introduction}

Building a large-scale quantum computer requires qubits that are able to retain their quantum coherence as well as a flexible architecture that enables both short and long range inter-qubit coupling schemes. Electrons confined to semiconductor double quantum dots (DQD)~\cite{Zwanenburg2013, Chatterjee2021, Burkard2023, Becher2023} possess both spin and charge degrees of freedom -- properties that are useful for coherent encoding of quantum information and fast manipulation by electric and magnetic fields. Spin coherence can be improved with material enrichment to remove spinful nuclear isotopes~\cite{Tyryshkin2012, Veldhorst2014, Eng2015} or with dynamical decoupling techniques~\cite{Alvarez2011, Muhonen2014}, and both spin and charge coherence can be enhanced at optimal points in parameter space~\cite{Koh2013, Reed2016, Thorgrimsson2017, Feng2021, Kratochwil2021, Pico-Cortes2021}. These make semiconductor quantum dots one of the most promising platforms for quantum computing.
 
However, inter-qubit coupling based on either Heisenberg exchange~\cite{Levy2002, DiVincenzo2000} or capacitive coupling~\cite{Neyens2019, Feng2021} is limited in range ($\sim$100's~nm and $\sim$1~$\mu$m, respectively),  posing a challenge to the scalability of quantum dot-based qubits. Long distance coupling schemes have been proposed that transfer qubit states across quantum dot arrays~\cite{Feng2018, Mills2019, Yoneda2021}, or make use of photons in microwave resonators to mediate qubit-qubit interactions using the approach of circuit quantum electrodynamics (cQED)~\cite{Childress2004, Burkard2006, Frey2012, Hu2012, Jin2012,  Benito2017, Tosi2017, Mi2017, Stockklauser2017,  Mi2018, Samkharadze2018, Landig2018, Scarlino2022, Warren2019, Burkard2020, Blais2021, Borjans2020, Harvey-Collard2022}. The cQED platform has the advantage of enabling a modular approach to scaling by linking well-characterized and locally optimized few-qubit modules~\cite{Vandersypen2017}.

In cQED, photons confined in superconducting microwave resonators can couple to both spin and charge degrees of freedom. Even though direct magnetic coupling of an electron spin to a photon is weak, the wavefunction of an electron can be delocalized in a DQD, creating a tunable electric dipole moment which can couple much more strongly to the electric field of the photon. In this way, the strong and tunable coupling of the electric dipole to the photon enables fast manipulation of a single electron DQD qubit, which may be operated as a spin or charge qubit depending on whether quantum information is encoded in the spin or charge degree of freedom. 

Leveraging the long coherence times for spins in silicon~\cite{Xue2022, Mills2022, Noiri2022}, strong coupling has  been demonstrated in DQD spin and charge qubits~\cite{Mi2017, Stockklauser2017, Mi2018, Samkharadze2018} and a triple QD spin qubit~\cite{Landig2018}. Coherent resonator-mediated interaction of two single-electron spin qubits in silicon has also been achieved~\cite{Borjans2020, Harvey-Collard2022}. Even though  spin qubits have longer coherence times than charge qubits, the bare charge-photon coupling $g_c$ can be an order of magnitude larger than the effective spin-photon coupling $g_s$~\cite{Samkharadze2018, Scarlino2022}, enabling much faster gates for the charge qubit. 
This enhanced coupling also opens up the possibility of studying photon-qubit interactions in the ultrastrong regime~\cite{Scarlino2022}. Even though the ultrastrong regime has been reached in superconducting platforms~\cite{Niemczyk2010}, it is a recent development for semiconductor quantum dot systems~\cite{Scarlino2022}.

The strong coupling regime is reached when coupling strengths exceed dissipation rates in the system, which are in general, charge and spin decoherence rates $\gamma_c$ and $\gamma_s$, and photon loss rate $\kappa$. Ultrastrong coupling, on the other hand, is achieved when coupling strengths become comparable to the bare system frequencies, which refer to the resonator frequency $\omega_r$ and the DQD charge and spin qubit transition frequencies, $\omega_c$ and $\omega_s$~\cite{Kockum2019, Forn-Diaz2019}. Conventionally, this is defined to be when the ratio of coupling strength to system frequency is above a threshold of 0.1~\cite{Kockum2019}. In this regime,  the rotating-wave approximation (RWA), which is a key approximation used in modelling light-matter interaction, breaks down. Due to the strict  requirement on low error rates for fault-tolerant quantum information processing, it is therefore necessary to go beyond RWA in order to accurately estimate error rates in quantum operations.


In this work, we theoretically investigate the two-qubit Controlled-Z (CZ) gate between a pair of DQD qubits coupled via microwave photons in a superconducting resonator. Detuning the qubit transition frequencies from the resonator frequency avoids entangling the qubits with the microwave photons. In this dispersive regime, the effective qubit-qubit interaction is weak ($\propto g^2/\Delta$), posing a challenge to rapid, high-fidelity operation~\cite{Warren2019}. Here, $g$ and $\Delta$  (without c/s subscripts) refer generically to either the charge or spin coupling to the photon, and the charge or spin qubit-resonator detuning. To implement faster quantum gates, one can parametrically drive the qubits -- by modulating their transition frequencies or coupling strengths at specific frequencies -- and induce sideband transitions with faster gate times $\propto 1/g$. 
Although quantum gates based on sideband transitions have been well-studied in superconducting qubits (see, e.g. Refs.~\cite{Wallraff2007, Blais2007, Beaudoin2012}), it has not been studied in detail for long-range manipulation of semiconductor quantum dot qubits.

Existing works~\cite{Abadillo-Uriel2021, Srinivasa2016, Srinivasa2024} have theoretically studied sideband transitions for DQD spin qubits using the RWA, while charge qubits have received less attention. Given the recent progress towards ultrastrong coupling, understanding the errors caused by the time-dependent terms that are ignored by RWA becomes critical for modeling high-fidelity operation.  In this manuscript, we go beyond RWA to show how CZ gates may be optimized for both DQD spin and charge qubits. 

The key results of our work can be summarized as follows.

First, we propose a novel parametric drive for the DQD charge qubit -- an essential step toward realizing a CZ gate. We demonstrate that this scheme is not only distinct from existing spin qubit protocols but also experimentally feasible under realistic bandwidth limitations.

Second, we move beyond the RWA, rigorously accounting for all error terms in both driven qubit systems. This comprehensive treatment uncovers previously hidden dependencies of gate fidelity on system and drive parameters, allowing us to map out high-fidelity regions in parameter space with much greater precision.

Third, our results show that the charge qubit delivers substantial performance advantages: it is markedly more resilient to both qubit dephasing and photon loss, achieves superior gate fidelities, and offers a broader, more robust operating window tolerant to variations in drive parameters.

The paper begins in Sec.~\ref{sec:Results} with an introduction of how the CZ gate can be implemented from sideband transitions, followed by a theoretical description of parametrically driven DQD charge and spin qubits. Resonant conditions and error terms ignored by the RWA are then explicitly derived. We then study the validity of the RWA from a heuristic analysis of the error terms. Gate fidelity is derived in terms of a perturbative expansion in the error terms and numerical results are calculated using the superoperator formalism which we detail in the Methods section. We consider the  parameter space of independent frequencies, and establish the boundaries of high-fidelity regions in the two-dimensional rescaled frequency space. We then analyze the fidelity of the exact gate without decoherence to quantify the effect of the error terms. This is followed by incorporating decoherence into the analysis of gate fidelity to establish the demands that fault-tolerance imposes on the decoherence rates. In Sec.~\ref{sec:Discussion}, we elaborate upon three main discussion points to validate our conclusions, namely bandwidth constraints on the charge qubit, further optimization of the spin qubit and an alternative control scheme. Finally, in Sec.~\ref{sec:Conclusion}, we evaluate the two qubit systems and discuss the outlook of implementing the CZ gate in current experimental systems.

\section{Theoretical Model}\label{sec:Results}


\subsection{Red-sideband transition and CZ gate}\label{sec:Red-sideband transition and CZ gate}

In this section, we describe the red-sideband Hamiltonian, followed by the protocol for achieving the CZ gate first reported in Ref.~\cite{Abadillo-Uriel2021}. In this work, we take $\hbar = 1$.

We consider one of the two qubits coupled to a single resonator mode at a time, in the dispersive regime where the qubit transition frequency $\omega_q$ is detuned from the resonator frequency $\omega_r$. We denote the qubit-resonator detuning by $\Delta \equiv \omega_r - \omega_q$. In this regime, leakage of photons from the resonator, known as Purcell decay, is strongly suppressed compared to the resonant case. Detuning reduces the effective interaction, leading to slower quantum gates, at rates that scale with $g^2/\Delta$. However, an external drive applied to the qubit (or the resonator) compensates for the reduced interaction, and can generate either red or blue sideband transitions~\cite{Blais2007, Srinivasa2016, Abadillo-Uriel2021}. These transitions are achieved under special resonant conditions on the drive frequency $\omega_d$ and amplitude $2\Omega$. The advantage of this so-called driven dispersive regime is that gate speeds scale with $g$, and are therefore much faster. 

The system described above can be represented in a Hilbert space $\mathbbm{H} = \mathbbm{H}_{\text{F}} \otimes \mathbbm{H}^{(1)}_{\text{q}} \otimes \mathbbm{H}^{(2)}_{\text{q}}$, where $\mathbbm{H}_{\text{F}}$ is the bosonic Fock space and $\mathbbm{H}^{(j)}_\text{q} = \mathbbm{C}^2$ denotes the qubit Hilbert space. The red-sideband Hamiltonian is
\begin{equation}\label{eq:red-sideband Hamiltonian}
	\HR^{(j)} (g, \phi) = \frac{ig}{2} (e^{i \phi}\a \p^{(j)} - e^{- i \phi} \c \m^{(j)}),
\end{equation}
where $\c$ ($\a$) is the photon creation (annihilation) operator and $\p^{(j)}$ ($\m^{(j)}$) is the qubit raising (lowering) operator acting on the $j$-th qubit, in anticipation of the two-qubit CZ gate which follows next. Note that we have implemented a canonical embedding map for all operators so that they act on the total Hilbert space. In the following, we exchangeably denote the composition of operators $\hat A$ and $\hat B$ by $\hat A \otimes \hat B$ or $\hat A \hat B$. It is a time-independent Hamiltonian that describes the exchange of excitations between a qubit and a photon mode in the resonator. Here, the phase $\phi$ of the drive is an additional tuning knob. Although there is no explicit dependence on drive frequency and amplitude, this is an idealization which generally arises from applying the RWA on real systems. 

The time-evolution generated by the red-sideband Hamiltonian is then
\begin{align}
    \SR^{(j)}(g \Delta t, \phi) = \exp \left [ -i\Delta t \HR^{(j)} (g,\phi) \right ],
\end{align}
which acts on the Fock space and the $j$-th qubit Hilbert space.

Following Ref.~\cite{Abadillo-Uriel2021}, composing a sequence of five red-sideband transitions with specific phases and gate times, the ideal CZ gate is given by
\begin{widetext}
\begin{align}
    \hat U_\text{cz} &= \hat M_0 \hat Z^{(1)} \left( \frac{\pi}{\sqrt{2}} \right) \hat Z^{(2)} \left(- \frac{\pi}{\sqrt{2}}\right) \SR^{(2)} \left(\pi, \pi \right) \SR^{(1)} \left(\frac{\pi}{2}, 0 \right) \SR^{(1)} \left(\sqrt{2} \pi, \frac{\pi}{2} \right) \SR^{(1)} \left(\frac{\pi}{2}, \pi \right) \SR^{(2)} \left(\pi, 0 \right)  \label{eq:Ucz0}\\
    &\equiv \hat M_0 \hat Z^{(1)}\left( \frac{\pi}{\sqrt{2}} \right) \hat Z^{(2)} \left(- \frac{\pi}{\sqrt{2}}\right) \SR(5) \SR(4) \SR(3) \SR(2) \SR(1),
    \label{eq:Ucz}
\end{align}
\end{widetext}
where the sequence of five qubit-photon interaction stages is followed by single-qubit $z$-rotations and a projection $\hat{M}_0 \equiv \ket{0}_r \bra{0}$ onto the zero-photon state of the resonator. In this sequence, when the $j$-th qubit is coupled to the resonator, the uncoupled qubit is tuned to a regime with a vanishing dipole moment. In the second line of the gate sequence, we have abbreviated the argument of each sideband to $\SR(b)$ to denote the evolution in the $b$-th stage, which automatically indicates the coupled qubit index $j_b$, the gate time $\Delta t_b$, and the phase $\phi_b$ in the $b$-th stage, as shown in Table~\ref{tab:abbreviation}. We do this to ease the notations for the calculations that come later. 

\begin{widetext}
\begin{center}
\begin{table}
\begin{tabular}{|c|c|c|c|c|c|}
\hline
stage & coupled qubit & duration  & phase & full notation & abbreviated  \\
$b$ & $j_b$ &  $g\Delta t_b$ &$\phi_b$  & $\SR^{(j_b)} \left(g\Delta t_b, \phi_b \right)$  & notation \\ \hline
1 & 2 & $\pi$ & 0 & $\SR^{(2)} \left(\pi, 0 \right)$   &   $\SR(1)$                 \\ \hline
2 & 1 & $\frac{\pi}{2}$  & $\pi$ & $\SR^{(1)} \left(\frac{\pi}{2}, \pi \right)$  & $\SR(2)$         \\ \hline
3 & 1 & $\sqrt{2}\pi$ & $\frac{\pi}{2}$ & $\SR^{(1)} \left(\sqrt{2} \pi, \frac{\pi}{2} \right)$ & $\SR(3)$  \\ \hline
4 & 1 & $\frac{\pi}{2}$ & 0 & $\SR^{(1)} \left(\frac{\pi}{2}, 0 \right)$  & $\SR(4)$          \\ \hline
5 & 2 & $\pi$ & $\pi$ & $\SR^{(2)} \left(\pi, \pi \right)$   & $\SR(5)$                  \\ \hline
\end{tabular}
\caption{The correspondence between the full and abbreviated notations in Eqs.~\eqref{eq:Ucz0} and~\eqref{eq:Ucz}.}\label{tab:abbreviation}
\end{table}
\end{center}
\end{widetext}

\subsection{Double quantum dot qubits}\label{sec:Double quantum dot qubits}

Our goal here is to describe the physical Hamiltonians for a pair of parametrically driven DQD qubits coupled to a resonator. The DQD qubits may be operated in the charge qubit or spin qubit regime, which we describe separately. Our focus will be to describe the external drive on each of these qubits. For the DQD spin qubit, this has been described in Refs.~\cite{Srinivasa2016, Abadillo-Uriel2021, Srinivasa2024}, so we only sketch the key ideas for completeness. More importantly, we explain our proposed, novel driving scheme for the DQD charge qubit.


\subsubsection{Physical Hamiltonians}

\begin{figure}[t]
	\includegraphics[width=1\linewidth]{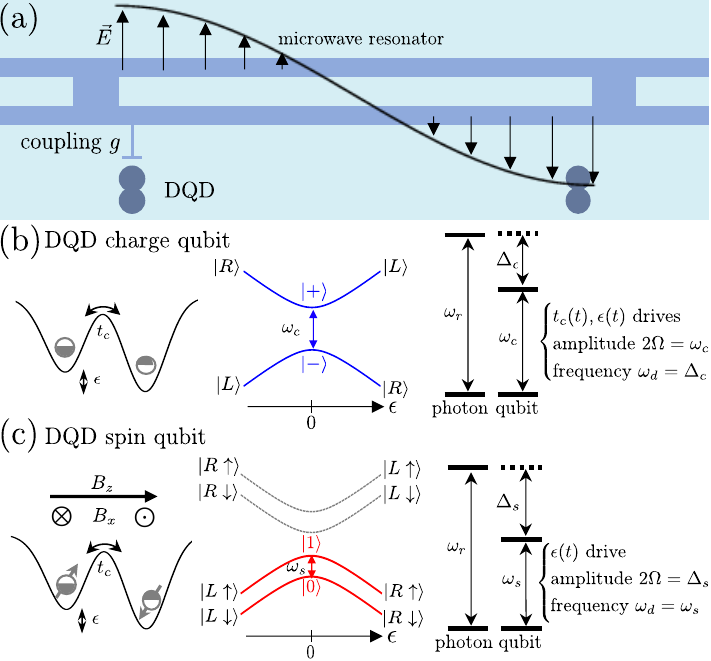}
	\caption{\label{fig:setup} (a) Schematic of two spatially separated DQD qubits in a circuit QED system. (b) Single-electron DQD charge qubit and energy scales. Tunnel coupling $t_c$ and orbital detuning $\epsilon$ control the charge qubit frequency $\omega_c$. Logical states (blue) are eigenstates of $\hat H_\text{DQD}$. Simultaneous  $t_c (t)$ and $\epsilon(t)$ drives with amplitude $2\Omega = \omega_c$ and drive frequency $\omega = \Delta$ enables red-sideband transitions. (c) Single-electron DQD spin qubit. The Zeeman $B_z$ field spin-splits each orbital; a spin-orbit field $B_x$ admixes spin and orbital states, enabling electrical control of the logical states (red). A resonant $\epsilon(t)$ drive with amplitude $2\Omega = \Delta$ enables red-sideband transitions. }\label{fig:F1}
\end{figure}

We begin by describing the undriven, physical Hamiltonians for a resonator and two DQD qubits, with the $j$-th qubit is coupled to the resonator (see Fig.~\ref{fig:setup}(a)). The Hamiltonians for the charge and spin qubits (denoted by subscripts `c', `s') in this scenario are given respectively by 
\begin{align}
	\hat H_\text{c}^{(j)} &=  \hat H_\text{res} + \sum_{q=1}^{2} \hat H_\text{DQD}^{(q)} + \hat H_\text{int}^{(j)}, \label{eq:charge qubit}\\
	\hat H_\text{s}^{(j)} &=  \hat H_\text{res}  + \sum_{q=1}^{2} \left( \hat H_\text{DQD}^{(q)} + \hat H_\text{B}^{(q)} \right) + \hat H_\text{int}^{(j)}. \label{eq:spin qubit physical}
\end{align}
Here, 
\begin{align}
	\hat H_\text{res} = \omega_r \c\a
\end{align}
is the resonator Hamiltonian. The DQD term comprises orbital detuning $\epsilon$ and tunnel coupling $t_c$ between the DQDs (see Fig.~\ref{fig:setup}(b, c)), and is given by
\begin{align}
	\hat H_\text{DQD}^{(q)} &= \frac{1}{2}(\epsilon^{(q)} \hat\tau_z^{(q)} + t_c^{(q)} \hat\tau_x^{(q)} ).
\end{align} 
Here $\tau_z^{(q)} \equiv \ket{L^{(q)}}\bra{L^{(q)}} - \ket{R^{(q)}}\bra{R^{(q)}}$ and $\tau_x^{(q)} \equiv \ket{L^{(q)}}\bra{R^{(q)}} + \ket{R^{(q)}}\bra{L^{(q)}}$, where $\ket{L^{(q)}}$ and $\ket{R^{(q)}}$ are the lowest orbitals in the left and right dots of the $q$-th DQD. The interaction term describes the bare coupling between the charge dipole of the coupled qubit and the electric field of the resonator, 
\begin{align}
	\hat H_\text{int}^{(j)} &= g_c^{(j)} \hat \tau_z^{(j)}  \left(\a + \c \right),
\end{align} 
where $g_c$ is the charge-photon coupling strength. 

The description of the spin qubit is similar to the charge qubit, except that there is additional magnetic term given by
\begin{align}\label{eq:spin-orbit term}
	\hat H_\text{B}^{(q)} = \frac{1}{2}(B_z^{(q)} \hat S_z^{(q)} + B_x^{(q)} \hat \tau_z^{(q)} \hat S_x^{(q)}).
\end{align} 
The first term describes an external, static Zeeman field. The second term describes a spin-orbit interaction which couples the spin and orbital degrees of freedom, which may be intrinsic to the material~\cite{Golovach2006, Weber2018, Krishnan2024, Aliyar2024} or engineered such that the left and right dots experience a transverse field gradient~\cite{Tosi2017, Hsueh2024}, e.g. with a micromagnet~\cite{Xian2014, Mi2017, Benito2017}. Here, $\hat S_z^{(q)} \equiv \ket{\uparrow^{(q)}}\bra{\uparrow^{(q)}} - \ket{\downarrow^{(q)}}\bra{\downarrow^{(q)}}$ and $\hat S_x^{(q)} \equiv \ket{\uparrow^{(q)}}\bra{\downarrow^{(q)}} + \ket{\downarrow^{(q)}}\bra{\uparrow^{(q)}}$, where $\ket{\updownarrow^{(q)}}$ are the electron spin components in the $z$-direction for the $q$-th DQD.

For both spin and charge qubits, $\epsilon = 0$ is an optimal operating point. At this sweet spot, the first derivative of the  qubit transition frequencies vanishes, enabling protection from charge noise in the $\epsilon$ parameter. Simultaneously, charge-photon coupling and, by extension, spin-photon coupling, are largest at this point. We take this as the working point for both qubits in this work.


\subsubsection{Parametric driving of DQD charge qubit}\label{sec:Parametric driving of DQD charge qubit}

In this section, we describe our proposed parametric drive on the DQD charge qubit. We denote the eigenbasis of $\hat H_\text{DQD}$ by the excited and ground orbital states $\{\ket{+}, \ket{-} \}$ (denoted by blue lines in Fig.~\ref{fig:setup}(b)). Expressing Eq.~\eqref{eq:charge qubit} in this basis, we have
\begin{align}\label{eq:charge_eigenbasis}
\hat H_\text{c}^{(j)} = &\omega_r \c\a + \sum_{q=1}^{2} \frac{\omega_c^{(q)}}{2} \hat\sigma_z^{(q)} \nonumber \\
&+ (\a + \c) (g_z^{(j)}\hat\sigma_z^{(j)} - g_x^{(j)} \hat\sigma_x^{(j)}) ,
\end{align}
where $\hat\sigma_z^{(q)} \equiv \ket{+^{(q)}}\bra{+^{(q)}} - \ket{-^{(q)}}\bra{-^{(q)}}$, $\hat\sigma_x^{(q)} \equiv \ket{+^{(q)}}\bra{-^{(q)}} + \ket{-^{(q)}}\bra{+^{(q)}}$, $g_z^{(j)} \equiv g_c^{(j)}\cos\vartheta^{(j)}$, $g_x^{(j)} \equiv g_c^{(j)}\sin\vartheta^{(j)}$, and $\vartheta^{(j)} = \arctan(t_c^{(j)} / \epsilon^{(j)})$ is the orbital mixing angle. The charge qubit splitting is 
\begin{align}
	\omega_c^{(q)} \equiv \sqrt{(t_c^{(q)})^{2} + (\epsilon^{(q)})^{2} }.\label{eq:charge splitting}
\end{align}

The DQD eigenstates are the logical states of the charge qubit, and are related to the left and right orbital states by
\begin{align}
    \ket{+} &= \cos(\tfrac{\vartheta}{2}) \ket L + \sin(\tfrac{\vartheta}{2}) \ket R,  \\
    \ket{-} &= - \sin(\tfrac{\vartheta}{2}) \ket L + \cos(\tfrac{\vartheta}{2}) \ket R. 
\end{align}
At the sweet spot, these are the symmetric and antisymmetric orbitals since $\cos(\tfrac{\vartheta}{2}) = \sin(\tfrac{\vartheta}{2}) = 1/\sqrt{2}$. The orbital detuning and tunnel coupling are controlled by electric fields applied to metal electrodes that define the DQD. Orbital detuning controls the relative energy difference between each QD and can be either positive or negative, but the sign of the tunnel coupling is non-negative, $t_c \ge 0$. We therefore choose to apply a sinusoidal drive with a finite offset, so that 
\begin{align}
    t_c(t) = & \Omega - \Omega \cos(\omega_d t + \phi) = 2\Omega  \sin (\frac{\omega_d t + \phi}{2})^2. \label{eq:t_c(t)}
\end{align} 
The drive frequency is $\omega_d$ and the drive phase is $\phi$. For ease of comparison with the spin qubit which we will see later, we define drive amplitude to be $2\Omega$. 

In addition, we require that the charge qubit splitting in Eq.~\eqref{eq:charge splitting} remains constant. Therefore, an additional drive to the orbital detuning must be applied concurrently, given by 
\begin{align}
    \epsilon(t) = 2\Omega \sqrt{1 - \sin(\frac{\omega_d t + \phi}{2})^4}.\label{eq:epsilon(t)}
\end{align} 
The qubit splitting and drive amplitude are thus both constant and equal to each other,  
\begin{align}
	\omega_c^{(j)} = 2\Omega.~\label{eq:charge qubit drive amplitude}
\end{align}

Finally, the drives lead to a time-dependent interaction of the charge qubit with the resonator. Assuming identical charge qubits, we drop the superscript $q$ on the qubit splitting term $\omega_c$, and express the driven charge qubit Hamiltonian as
\begin{align} \label{eq:charge qubit driven}
\hat H_\text{c}^{(j)}(t) = &\omega_r \c\a + \sum_{q = 1}^2 \frac{\omega_c}{2} \Z^{(q)} \nonumber \\
&+ (\a + \c) \left(g_z^{(j)}(t) \Z^{(j)} - g_x^{(j)}(t) \X^{(j)} \right), 
\end{align}
where 
\begin{align}
	g_x^{(j)}(t) &=  g_c^{(j)}  \sin (\frac{\omega_d t + \phi}{2})^2,~\label{eq:gx(t)} \\
	g_z^{(j)}(t) &=  g_c^{(j)} \sqrt{1 - \sin(\frac{\omega_d t + \phi}{2})^4}.~\label{eq:gz(t)}
\end{align}

We note that Eq.~\eqref{eq:charge qubit driven} is written in terms of the eigenstates of $\hat H_\text{DQD}$ at the sweet spot, so the drive should be treated perturbatively. This parametric drive is intended to induce Rabi oscillations between the logical qubit states, described by the $\X$ term in Eq.~\eqref{eq:charge qubit driven}. An alternative to the drive described in Eq.~\eqref{eq:t_c(t)} could be to reduce the magnitude of the oscillatory term so that it is smaller than the constant offset. This would also satisfy the condition of non-negative tunnel coupling. However, this choice leads to a smaller value of $g_x$ and produces a smaller Rabi frequency and longer gate time. Our choice, as described in Eq.~\eqref{eq:t_c(t)}, is thus optimal.


\subsubsection{Parametric driving of DQD spin qubit}
\label{sec:Parametric driving of DQD spin qubit}

Now, we describe the driven Hamiltonian for the DQD spin qubit, which follows the protocol described in Ref.~\cite{Srinivasa2024}. Unlike Ref.~\cite{Srinivasa2024} however, we do not assume the RWA, and thus all time-dependences of the exact Hamiltonian can be seen when we show how the red-sideband Hamiltonian is induced in the following section.

For the spin qubit, when the orbital spacing is larger than the Zeeman splitting $\sqrt{\epsilon^2 + t_c^2} > B_z$, the ground manifold comprises mainly the $\ket{- \uparrow}$ and $\ket{- \downarrow}$ spin states (denoted by red lines in Fig.~\ref{fig:setup}(c)) with some admixture of excited orbital states. States with the same spin but different orbitals are coupled by the electric field of the resonator through their dipole moments. The combination of these two effects couples the two spin states in the ground manifold.

The spin qubit may be driven by a sinusoidal signal with amplitude $\epsilon_0$ applied to the orbital detuning at the sweet spot, $\epsilon^{(j)}(t) = - \epsilon_0^{(j)} \cos(\omega_d t + \phi)$. The drive amplitude should be sufficiently small so that the qubit energy splitting remains approximately constant. In the rotating frame defined by $e^{i (\pi/4) \hat \tau_y^{(j)}}$, this appears as a transverse drive, i.e. a rotation around the $x$-axis in the rotated frame.

Next, with a weak magnetic field gradient, the spin-orbit term in Eq.~\eqref{eq:spin-orbit term} may be treated perturbatively. This allows a Schrieffer-Wolff transformation to be applied such that the low-energy dynamics are decoupled and described by an effective two-level Hamiltonian~\cite{Benito2017, Srinivasa2024}. Similar to charge qubit case, we also assume identical spin qubits and drop the superscript $q$ on the qubit splitting term $\omega_s$. The effective Hamiltonian of the driven spin qubit thus reads
\begin{align}\label{eq:spin qubit driven}
	\hat H_\text{s,eff}^{(j)}(t) = &\omega_r \c\a +  \sum_{q=1}^{2} \frac{\omega_s}{2} \tilde\sigma_z^{(q)} -  2 \Omega \cos(\omega_d t + \phi) \tilde\sigma_x^{(j)} \nonumber \\
	+ & g_s^{(j)} (\a + \c) \tilde\sigma_x^{(j)}.
\end{align}
Here, we use a slightly different Pauli operator notation from the charge qubit case, $\tilde\sigma$ with a tilde instead of a hat. This is to indicate the equivalence in the two-level descriptions, but with the understanding that for the spin qubit, the qubit operators are written in the basis of the lowest two eigenstates, 
\begin{align}
    \ket {1} \approx \ket{- \uparrow},~
    \ket {0} \approx \ket {- \downarrow},
\end{align}
shown as red lines in the eigenspectrum in Fig.~\ref{fig:setup}(c).
The new Pauli operators are $ \tilde\sigma_z^{(q)} \equiv \ket{1^{(q)}}\bra{1^{(q)}} - \ket{0^{(q)}}\bra{0^{(q)}}$ and $ \tilde\sigma_x^{(q)} \equiv \ket{1^{(q)}}\bra{0^{(q)}} + \ket{0^{(q)}}\bra{1^{(q)}}$. The spin qubit splitting scales with the Zeeman field $\omega_s \approx B_z$. Spin-photon coupling is $g_s^{(j)} \equiv g_c^{(j)} |\sin (\Phi^{(j)}/2)|$, which is tunable by controlling tunnel coupling or spin-orbit coupling. The effective drive amplitude is $2 \Omega \equiv \epsilon_0 |\sin (\Phi^{(j)}/2)|$, where $\Phi \equiv \arctan\frac{B_x}{t_c-B_z}$ is the spin-orbit mixing angle.


\subsection{Exact Hamiltonians in rotating frames}\label{sec:Exact Hamiltonians in rotating frames}

Now, we are ready to show that the red-sideband Hamiltonian can be induced with appropriate resonant conditions on the drives applied to the charge and spin qubits. To see this, we will apply unitary transformations $\hat R_{c(s)}$ to go into rotating frames using $\hat R_\text{c(s)}^\dag(\hat H_\text{c(s)} - i \partial_t) \hat R_\text{c(s)}$. We note that unlike earlier works that make use of the RWA, we keep all time-dependent terms explicitly in our derivation.

We first transform into the frame rotating with the bare qubit and resonator Hamiltonians with the unitaries
\begin{align}
	\hat R_\text{1,c}(t) &=\exp{ -i t \left(\omega_r \c\a + \sum_{q = 1}^2 \frac{\omega_{c}}{2} \hat { {\sigma}}_z^{(q)} \right)}, \label{eq:R_{1,c}}\\
	\hat R_\text{1,s}(t) &= \exp{ -i t \left(\omega_r \c\a + \sum_{q = 1}^2 \frac{\omega_{s}}{2}  {\tilde {\sigma}}_z^{(q)}\right)},\label{eq:R_{1,s}}
\end{align}
on the driven charge and spin qubit Hamiltonians (Eqs.~\eqref{eq:charge qubit driven},~\eqref{eq:spin qubit driven}). In the first rotating frame (denoted by subscript ``rf1''), the Hamiltonians are
\begin{align}
	\hat H_\text{c,rf1}^{(j)}(t) 
	= &  (\a e^{-i\omega_r t} + \c e^{i\omega_r t}) \nonumber \\ 
	\otimes &\left[ g_z^{(j)}(t) \Z^{(j)} - g_x(t) \left( e^{i \omega_c t } \hat \sigma_+^{(j)} + e^{-i \omega_c t } \hat \sigma_-^{(j)} \right)\right],~\label{eq:Hc1} \\
	\hat H_\text{s,rf1}^{(j)}(t) = &(\a e^{-i\omega_r t} + \c e^{i\omega_r t}) \left( g_s^{(j)} - 2 \Omega \cos(\omega_d t + \phi) \right)  \nonumber \\
	\otimes &\left( e^{i\omega_s t}\tilde\sigma_+^{(j)} + e^{-i \omega_s t} \tilde\sigma_-^{(j)}\right).
\end{align}
Next, we move to a second rotating frame defined by a fixed rotation about the $z$-axis, 
\begin{align}
	\hat R_\text{2,c} &= \exp{ - i\frac{\pi}{4} \sum_q \Z^{(q)}}, \\
	\hat R_\text{2,s}(\phi) &= \exp{- i \left(\frac{\phi}{2} + \frac{\pi}{4}\right) \sum_q\tilde \sigma_z^{(q)}}.
\end{align}
In this frame, the Hamiltonians are
\begin{align}
	\hat H_\text{c,rf2}^{(j)}(t) 
	= & (\a e^{-i\omega_r t} + \c e^{i\omega_r t}) \nonumber \\ 
	\otimes &\left[g_z^{(j)}(t) \Z^{(j)} - i g_x(t) \left( e^{i \omega_c t } \hat \sigma_+^{(j)} - e^{-i \omega_c t } \hat \sigma_-^{(j)} \right) \right],~\label{eq:charge qubit rf} \\
	\hat H_\text{s,rf2}^{(j)}(t) = & (\a e^{-i\omega_r t} + \c e^{i\omega_r t}) \left(i g_s^{(j)} - 2 i \Omega \cos(\omega_d t + \phi) \right) \nonumber \\ 
	\otimes &\left( e^{i(\omega_s t + \phi)}\tilde\sigma_+^{(j)} - e^{-i (\omega_s t + \phi)} \tilde\sigma_-^{(j)}\right).
\end{align}
We are done with the transformations for the charge qubit, but the spin qubit requires two more steps. For the spin qubit, we  rotate with the Rabi drive next, given by
\begin{align}
	\hat R_\text{3,s} (t) = & \exp{- i (\Omega t) \tilde \sigma_y^{(j)}}. \label{eq:R_{3,s}}
\end{align}
This leads to the third rotating frame Hamiltonian,
\begin{widetext}
\begin{align}
	\hat H_\text{s,rf3}^{(j)}(t) 
	= &\left( \a e^{-i\omega_r t} + \c e^{i\omega_r t}\right)
	\left[ 2 \Omega  \left( \sin(\omega_s t + \phi)\cos(\omega_d t + \phi) \cos(2\Omega t)\tilde \sigma_x^{(j)} + \left(\cos(\omega_s t + \phi)\cos(\omega_d t + \phi) -\tfrac{1}{2} \right)  \tilde \sigma_y^{(j)} \right. \right. \nonumber \\
	& \left . ~~~~~~+  \sin(\omega_s t + \phi)\cos(\omega_d t + \phi)  \sin(2\Omega t) \tilde \sigma_z^{(j)} \right) \nonumber  \\ 
	& \left. - g_s \left( \sin(\omega_s t + \phi) \cos(2\Omega t) \tilde \sigma_x^{(j)}  + \cos(\omega_s t + \phi) \tilde \sigma_y^{(j)} + \sin(\omega_s t + \phi)  \sin(2\Omega t) \tilde \sigma_z^{(j)} \right)  \right ].
\end{align}
\end{widetext}

The spin-photon interaction ($g_s$ term) contains a circular  drive whose axis rotates at a frequency $2\Omega$ on the $xz$-plane. A final transformation is required to align the axis of the circular drive  to the $z$-direction, given by
\begin{align}
	\hat R_\text{4,s} = & \exp{ i (\pi/4)\sum_q \tilde \sigma_x^{(q)}}. \label{eq:R_{4,s}}
\end{align}
The driven spin qubit Hamiltonian in the final rotating frame (``rf4'') reads as
\begin{widetext}
\begin{align}
	\hat H_\text{s,rf4}^{(j)}(t) =&\left( \a e^{-i\omega_r t} + \c e^{i\omega_r t}\right) 
	\left[ - g_s  \left(e^{i2\Omega t} \sin(\omega_s t + \phi) \tilde \sigma_+^{(j)} 
	 + e^{-i2\Omega t} \sin(\omega_s t + \phi)  \tilde \sigma_-^{(j)} 
	 + \cos(\omega_s t + \phi)  \tilde \sigma_z^{(j)} \right) \right.  \nonumber \\
	&  ~~~~~~~~~~~~~~~~~~~~~~~~~~+2 \Omega \left ( \cos(2\Omega t) \sin(\omega_s t + \phi) \cos(\omega_d t + \phi) \tilde \sigma_x^{(j)} 
	 - \sin(2\Omega t) \sin(\omega_s t + \phi) \cos(\omega_d t + \phi)\tilde \sigma_y^{(j)} \right. \nonumber \\
	& \left. \left. ~~~~~~~~~~~~~~~~~~~~~~~~~~~~~~~~~ + \left(\cos(\omega_s t + \phi)\cos(\omega_d t + \phi) - \tfrac{1}{2}\right)\tilde \sigma_z^{(j)}  \right ) \right]. \label{eq:spin qubit rf}
\end{align}
\end{widetext}

We note that no approximations have been made and all rotating frame Hamiltonians are exact. Although the full expressions are rather cumbersome, the purpose of doing so is to deduce the resonance conditions for the red-sideband transition and, more importantly, analyse all the error terms that the RWA neglects.


\subsection{Resonance conditions, rotating-wave approximation and error terms}\label{sec:Resonance conditions, rotating-wave approximation and error terms}

We are now in a position to derive the resonant conditions that induce the red-sideband from the final rotating frame Hamiltonians of the driven DQD charge and spin qubits (Eqs.~\eqref{eq:charge qubit rf} and~\eqref{eq:spin qubit rf}). First, we define the resonator-qubit detuning for the charge (spin) qubits: $\Delta_{c(s)} = \omega_r - \omega_{c(s)}$.


\subsubsection{Resonance conditions for DQD charge qubit}

For the charge qubit, we observe that the identity $ \sin (\frac{\omega_d t + \phi}{2})^2= \left( e^{i\omega_d t}e^{i \phi} + e^{-i\omega_d t}e^{-i \phi} - 2 \right)$ allows us to  re-express the $g_x(t)$ term of the driven charge qubit. We write the driven charge qubit Hamiltonian in its final rotating frame (``rf2'') in a more instructive way, as
\begin{widetext}
\begin{align}
	\hat H_\text{c,rf2}^{(j)}(t) 
	=& i \frac{g_c^{(j)}}{4} \left( \a \pj e^{i \phi} e^{-i (\Delta_c-\omega_d) t} - \c \mj e^{-i \phi}  e^{i (\Delta_c- \omega_d) t} \right)  \nonumber \\
	& + i \frac{g_c^{(j)}}{4} \left[ \a \pj \left( e^{-i \phi} e^{-i (\Delta_c+\omega_d) t} -2  e^{-i \Delta_c t} \right)- \c \mj \left( e^{i \phi}  e^{i (\Delta_c+\omega_d) t} -2 e^{i \Delta_c t} \right) \right] \nonumber \\
	& - i g_c^{(j)}  \left( e^{i\omega_d t}e^{i \phi} + e^{-i\omega_d t}e^{-i \phi} - 2 \right) \left(\c \pj e^{i (\omega_c+\omega_r) t} - \a \mj e^{-i (\omega_c+\omega_r) t}   \right)  \nonumber \\ 
	& + g_z^{(j)}(t) \left(\a e^{-i\omega_r t} + \c e^{i\omega_r t}\right) \Zj .~\label{eq:charge qubit resonance}
\end{align}
\end{widetext}
By inspection of the terms within the first set of brackets in Eq.~\eqref{eq:charge qubit resonance}, we deduce that when $\omega_d = \Delta_c$, the first line becomes time-independent and identical to the red-sideband Hamiltonian with a coupling strength of $g = g_c/2$. Together with the drive amplitude condition in Eq.~\eqref{eq:charge qubit drive amplitude}, the resonance conditions for the charge qubit are
\begin{align}\label{eq:charge_resonance}
\text{charge qubit resonance: }
	\begin{cases}
		\omega_d =  \Delta_c, \\
		2\Omega =  \omega_c.
	\end{cases}
\end{align}
This is denoted in Fig.~\ref{fig:setup}(b). We note that the drive frequency is equal to the resonator-qubit detuning instead of the qubit frequency. Finally, the resonantly driven charge qubit in the final rotating frame is described exactly by the Hamiltonian 
\begin{widetext}
\begin{align}
	&\hat H_\text{c,exact}^{(j)} \left(\frac{g_c}{2}, t \right) 
	=  i \frac{g_c^{(j)}}{4} \left( \a \pj e^{i \phi} - \c \mj e^{-i \phi} \right) + {\hat V_\text{c,red}^{(j)}(t)}  + {\hat V_\text{c,blue}^{(j)}(t)} + {\hat V_\text{c,long}^{(j)}(t)}, \label{eq:exact H_charge} \\
& \text{where } 
\begin{cases}\label{eq:charge_errors}
	{\hat V_\text{c,red}^{(j)}(t)} =  { i \frac{g_c^{(j)}}{4} \left( \a \pj \left(e^{-i \phi} e^{-2i\omega_d t} - 2e^{-i \omega_d t}\right)- \c \mj \left(e^{i \phi}  e^{2i \omega_d t} - 2 e^{i \omega_d t}\right) \right)}, \\
	{\hat V_\text{c,blue}^{(j)}(t)} =  { i g_c^{(j)}  \sin (\frac{\omega_d t + \phi}{2})^2 \left( \a \mj e^{-i (4\Omega+\omega_d) t} - \c \pj e^{i (4\Omega+\omega_d) t}   \right)}, \\
	{\hat V_\text{c,long}^{(j)}(t)} =  {g_c^{(j)} \sqrt{1 - \sin(\frac{\omega_d t + \phi}{2})^4}\left(\a e^{-i(2\Omega +\omega_d) t} + \c e^{i (2\Omega +\omega_d) t}\right)} \Zj .
\end{cases}
\end{align}
\end{widetext}
We can see that the time-independent term is the red-sideband Hamiltonian, which is our goal. However, the evolution caused by the time-dependent terms are undesirable and lead to inaccuracies, and we label them error terms.

Now, we will explain the physical meaning of the error terms. Terms $\hat V_\text{c,red}^{(j)}(t)$ and $\hat V_\text{c,blue}^{(j)}(t)$ arise from counter-rotating terms in the red-sideband and blue-sideband respectively. While red-sidebands induce an exchange of excitations between the qubit and the resonator ($\hat\sigma_+ \a$ and $\hat\sigma_- \c$ terms), blue-sidebands correspond to simultaneous excitations and emissions ($\hat\sigma_+ \c$ and $\hat\sigma_- \a$ terms). We note that a blue-sideband can be induced by a different resonant condition, but it is a slower gate~\cite{Abadillo-Uriel2021} and we do not consider it in this work.

The final error term $V_\text{c,long}^{(j)}(t)$ arises from the longitudinal precession of the qubit when there are unequal orbital admixtures in the instantaneous eigenstates. This effect is due to the drive on orbital detuning $\epsilon(t)$ in Eq.~\eqref{eq:epsilon(t)}, which oscillates asymmetrically away from and back toward the sweet spot. The longitudinal error displaces the mode of the photon at the cost of acquiring a phase on the qubit. 

Since all the error terms oscillate rapidly, the usual RWA assumes that they average to zero and have negligible effect on system dynamics. Using the RWA, we get
\begin{align}
	\hat H_\text{c,exact}^{(j)}  &\approxRWA   i \frac{g_c^{(j)}}{4} \left( \a \pj e^{i \phi} - \c \mj e^{-i \phi} \right) 
	 = \HR^{(j)} \left(\tfrac{g_c^{(j)}}{2}, \phi \right). ~\label{eq:charge qubit red-sideband approximation}
\end{align}
As a consequence, there is no dependence on drive amplitude and drive frequency in the dynamics governed by the approximation in Eq.~\eqref{eq:charge qubit red-sideband approximation}. Therefore, any limitations on the tunability of system parameters and optimal gate fidelity cannot be derived within the RWA. 


\subsubsection{Resonance conditions for DQD spin qubit}\label{sec:resonance spin qubit}

We now analyse the final rotating frame Hamiltonian for the spin qubit to deduce its resonance conditions. From the first line of Eq.~\eqref{eq:spin qubit rf}, we observe that in order to produce red-sideband terms, we require
\begin{align}\label{eq:spin_resonance}
\text{spin qubit resonance: }
	\begin{cases}
		\omega_d =  \omega_s, \\
		2\Omega =  \Delta_s.
	\end{cases}
\end{align}
This is illustrated in Fig. 1(c). Thus, the resonantly driven spin qubit Hamiltonian in the final rotating frame is given by the exact Hamiltonian
\begin{widetext}
\begin{align}
	&\hat H_\text{s,exact}^{(j)}(g_s, t) 
	=  i \frac{g_s^{(j)}}{2} \left(  \a \tilde\sigma_+^{(j)} e^{i \phi} - \c \tilde\sigma_-^{(j)} e^{-i \phi} \right) + \hat V_\text{s,red}^{(j)}(t) + \hat V_\text{s,blue}^{(j)}(t) + \hat V_\text{s,long}^{(j)}(t) + \hat V_\text{s,drive}^{(j)}(t), \label{eq:exact H_spin}\\
& \text{where } 
\begin{cases}\label{eq:spin_errors}
	\hat V_\text{s,red}^{(j)}(t) =   i \frac{g_s^{(j)}}{2} \left( - \a \tilde\sigma_+^{(j)} e^{-i2\omega_d t - i \phi}  + \c \tilde\sigma_-^{(j)} e^{i \phi}  e^{i2\omega_d t + i \phi}  \right), \\
	\hat V_\text{s,blue}^{(j)}(t) =  - g_s^{(j)}  \sin (\omega_d t + \phi) \left( \a \tilde\sigma_-^{(j)} e^{-i (4\Omega+\omega_d) t} + \c \tilde\sigma_+^{(j)} e^{i (4\Omega+\omega_d) t }   \right), \\
	\hat V_\text{s,long}^{(j)}(t) =  - g_s^{(j)}  \cos (\omega_d t + \phi)\left( \a e^{-i (2\Omega+\omega_d) t} + \c e^{i (2\Omega+\omega_d) t }  \right)  \tilde\sigma_z^{(j)} , \\
	\hat V_\text{s,drive}^{(j)}(t) = \Omega e^{i2\Omega t} \sin(2\omega_d t + 2 \phi) \tilde\sigma_+^{(j)} + \Omega e^{-i2\Omega t} \sin(2\omega_d t + 2 \phi) \tilde\sigma_-^{(j)} + \Omega \cos(2\omega_d t + 2 \phi) \tilde\sigma_z^{(j)}.
\end{cases}
\end{align}
\end{widetext}

Error terms $\hat V_\text{s,red}^{(j)}(t), \hat V_\text{s,blue}^{(j)}(t)$ and $\hat V_\text{s,long}^{(j)}(t)$ have the same physical meaning as those for the charge qubit, because arise in the same manner, and we do not repeat it here. Instead, we note the presence of an additional drive error term $\hat V_\text{s,drive}^{(j)}(t)$ that is absent for the charge qubit. This error term originates from non-resonant frequencies generated by the sinusoidal modulation of the orbital detuning. These frequencies drive spurious dynamics due to precessions around a time-dependent axis. This motion can be decomposed into two distinct components: (i) a primary precession about an axis counter-rotating uniformly in the equatorial plane at a rate $2\Omega$, and (ii) a secondary precession about a tilted axis, orthogonal to the plane defined by the primary axis and the $z$-axis, oscillating at a rate $2\omega_d$. The interplay between these two precessions creates a nonlinear, time-varying trajectory, leading to complex and undesirable dynamics that deviate significantly from the intended Rabi oscillation behavior. Since the drive error has a magnitude of $\Omega$, it becomes significant under strong driving when drive amplitude is comparable to system frequencies. In the simpler case of a strongly driven qubit (without coupling to a resonator), it is known that strong driving gives rise to a shift in qubit frequency -- termed the Bloch-Siegert shift --  and can be accounted for by using gate-time corrections to improve gate fidelity~\cite{Lu2012}. We will explain, in Sec.~\ref{sec:Discussion} that for our case, these gate-time corrections do not improve CZ gate fidelity.

Similar to the charge qubit case, after applying the RWA, we get
\begin{align}
	\hat H_\text{s,exact}^{(j)} &\approxRWA   i \frac{g_s^{(j)}}{2} \left( \a \pj e^{i \phi} -\c \mj e^{-i \phi} \right) 
	 = \HR^{(j)} \left(g_s^{(j)}, \phi \right). ~\label{eq:spin qubit red-sideband approximation}
\end{align}
The important differences are the coupling strengths $g = g_s^{(j)}$ for the spin qubit vs $g = g_c^{(j)}/2$ for the charge qubit, the resonance conditions (Eqs.~\eqref{eq:charge_resonance},~\eqref{eq:spin_resonance}) and the details of the error terms.


\subsection{RWA validity conditions and Fourier decomposition of error terms}\label{sec:RWA validity conditions and Fourier decomposition of error terms}

In this section, we first present a heuristic argument for the condition for the validity of the RWA, before we analyze the Fourier components of the error terms and pinpoint the occurrence of spurious resonances.


\subsubsection{General RWA consideration}

We saw in Section~\ref{sec:Resonance conditions, rotating-wave approximation and error terms} that parametrically driving the charge and spin qubits induces the red-sideband Hamiltonian, and all the error terms are neglected by the RWA.  The approximation assumes that  oscillations are rapid and average to zero. As such, if a particular system or drive frequency causes an error term to oscillate slowly or become time-independent, the RWA no longer holds. Such spurious resonances will thus cause inaccuracies in gate operations even as the parametric drive is constrained by the sideband resonance conditions. 

We first explain how the RWA conditions can be deduced from the Fourier decomposition of the error terms. The RWA can be understood by looking at the Dyson series expansion of the time-evolution operator in the interaction picture,
\begin{align}\label{eq:Dyson_series}
	\hat{U}(t)=&  \mathbbm{\hat 1} + \left(-i \right)\int_0^t{dt_1 \hat V(t_1)} \nonumber \\
	+ &\left(-i \right)^2  \int_0^t{dt_1} \int_0^{t_1}{dt_2  \hat V(t_1) \hat V(t_2)} + \dots 
\end{align}
If we express the perturbation (which in our case would be the error terms) in the frequency domain as $\hat V(t) = \sum_j \lambda_j e^{i \omega_j t} \hat{K}_j$ where $\omega_j$ is the frequency, $\lambda_j$ is the coupling strength  and $\hat K_j$ are the system's operators,  it becomes clear that the $n$-th order integral in the series yields terms with magnitude $\left(\lambda_j / \omega_j \right)^n$. Essentially, for the RWA to work, the coupling strength has to be much smaller than the oscillation frequency and the evolution occurs over a timescale that is much longer than that set by the oscillation frequencies, meaning $\abs{ \lambda_j / \omega_j }\ll 1$. This ensures that the fast-oscillating terms don’t significantly impact the system’s evolution.


\subsubsection{Dimensionless quantities}

Before we continue, we introduce dimensionless quantities (denoted with a tilde) which will be convenient for analyses. We rescale all frequencies by the coupling strength $g$, where $g = g_c/2$ for the charge qubit and $g = g_s$ for the spin qubit, and introduce dimensionless frequencies and time as follows.
\begin{align}
	\tilde\omega_r \equiv \frac{\omega_r}{g}, ~ \tilde\omega_{c/s} \equiv \frac{\omega_{c/s}}{g}, ~ \tilde\Omega \equiv \frac{\Omega}{g},  ~\tilde\omega_d \equiv \frac{\omega_d}{g}, ~ \Delta \tau \equiv  g \Delta t. 
\end{align}

We note that there are five frequencies in the driven system -- one each from the resonator and qubit, two from the drive and one from the qubit-resonator coupling. With two resonance constraints, three independent parameters remain. We choose these to be resonator frequency $\omega_r$,  coupling strength $g$ and drive amplitude $2\Omega$. By working with dimensionless quantities, this three-dimensional parameter space is reduced, allowing us to consider a smaller two-dimensional space of dimensionless frequencies $\{2\tilde\Omega, \tilde \omega_r\}$ without loss of generality. 

Our selection of independent variables is intentional: we focus on the system, coupling, and drive frequencies. While none of these parameters are unique in principle, experimentally, the frequency of a resonator is typically fixed and lacks tunability. It is thus useful to understand, {\em a priori}, which resonator frequencies yield the best fidelities during the experiment design stage, which is why we chose resonator frequency as an independent variable. The validity of the RWA conditions is directly influenced by drive amplitude as we shall see, making it another logical choice.  Additionally, coupling strengths vary significantly between devices and even working points, making results that highlight the dependence on rescaled frequencies (and consequently on coupling strength) valuable.


\subsubsection{Fourier decomposition and RWA conditions of DQD charge qubit}\label{sec:RWA_charge_qubit}

Let us now examine the Fourier composition of each of the three charge qubit error terms in Eqs.~\eqref{eq:charge_errors}.  The red- and blue-sideband error terms $\hat V_\text{c,red}^{(j)}(t)$ and $\hat V_\text{c,blue}^{(j)}(t)$ contains the product of operators with complex exponentials and sinusoidal functions. Their frequencies can therefore be listed easily by inspection. 

The term $\hat V_\text{c,red}^{(j)}(t)$ has coupling strength of order $g_c/4$ and contains frequencies $\omega \in \{\pm \omega_d, \pm 2\omega_d \}$. This means that the drive frequency on the charge qubit cannot be too small. Since the resonance condition is $\omega_d = \Delta_c \equiv \omega_r - \omega_c$, this implies that a large resonator-qubit detuning $ \Delta_c$ is advantageous. For RWA to be valid, 
\begin{align}\label{eq:charge_RWA_upper}
	\abs{\frac{g_c/4}{\Delta_c}} = \abs{\frac{1}{2(\tilde\omega_r - 2\tilde\Omega)}} \ll 1,
\end{align}
which represents a constraint on the rescaled drive amplitude, upper bounded by the rescaled resonator frequency.

Next, the term $\hat V_\text{c,blue}^{(j)}(t)$ has coupling strength of order $g_c$  and contains frequencies $\omega \in \{\pm 4\Omega, \pm(4\Omega + \omega_d), \pm(4\Omega + 2\omega_d) \}$. Spurious resonances occur when these frequencies approach zero. Expressed in terms of resonator and qubit frequencies, these are $\{\pm 2\omega_c, \pm(\omega_c + \omega_r), \pm 2\omega_r \}$. Since the charge qubit and resonator frequencies are in the microwave range and non-vanishing, this error term does not contain spurious resonances in practice. Therefore, the RWA is valid when 
\begin{gather} \label{eq:charge_RWA_lower}
	\abs{\frac{g_c}{2\omega_c}} = \abs{\frac{1}{2\tilde \Omega}}  \ll 1, ~
	\abs{\frac{g_c}{\omega_c + \omega_r}} = \abs{\frac{2}{\tilde\omega_c + \tilde\omega_r}}  \ll 1, \nonumber \\
	\abs{\frac{g_c}{2\omega_r}} = \abs{\frac{1}{\tilde\omega_r}} \ll 1.
\end{gather} 

These RWA conditions place a lower limit on drive amplitude and system frequencies to be much larger than the coupling strength. The first RWA condition puts a lower limit on drive amplitude. In the ultrastrong coupling regime when coupling strength $g_c$ becomes a significant fraction of either system frequency $\omega_c$ or $\omega_r$,  the above RWA conditions break down and one would have to take into account the error terms for an accurate prediction of the dynamics. This is a relevant scenario given recent experiments which report ultrastrong coupling of a resonator-DQD charge qubit device, where $\abs{g_c/\omega_c} \approx \abs{g_c/\omega_r} > 0.1$~\cite{Scarlino2022}.

The longitudinal error term features a complicated time-dependent pre-factor that arises from the orbital detuning drive. To decompose the $\epsilon(t)$ drive into complex exponentials, we utilize the generalized binomial expansion, yielding
\begin{align} \label{eq:epsilon_expansion}
    \epsilon(t) &= \sqrt{1 - \sin(\frac{\omega_d t + \phi}{2})^4} \nonumber \\
	&= \sum_{k=0}^{\infty} \sum_{m=0}^{4k} \frac{C^{1/2}_{k}  C^{4k}_{m} \left(-1\right)^{m}}{ \left(2i \right)^{4k} } e^{i (2k - m) (\omega_d t + \phi)} \nonumber \\
	&\equiv \sum_{k=0}^{\infty} \sum_{m=0}^{4k} d_{km} e^{i (2k - m) (\omega_d t + \phi)}, 
\end{align}
where in the last line above,  we have defined the coefficient $d_{km} \equiv  \frac{C^{1/2}_{k}  C^{4k}_{m} \left(-1\right)^{m}}{ \left(2i \right)^{4k} }$. The binomial coefficient is easily computed from integer factorials, as $C^{4k}_{m} = \frac{(4k)!}{m! (4k-m)!}$, while the generalized binomial coefficient is more easily computed in terms of the gamma functions, $C^{1/2}_{k} =  \frac{\Gamma(3/2)}{\Gamma(k+1)\Gamma(3/2-k)}$. Note that we give an alternative analytical expression for Eq.~\eqref{eq:epsilon_expansion} in Appendix~\ref{supp:detuning drive}.
 
We can now examine the longitudinal error term for its Fourier components. It includes the sum of a pair of frequencies $\pm(2\Omega + \omega_d)$ arising from the longitudinal coupling between the resonator and the charge eigenstates, as well as an infinite set of frequencies $(2k - m)\omega_d$ stemming from the orbital detuning drive. The former may be thought of as a coupling-dependent shift of the qubit frequency as it originates from the $\Z (\a + \c)$ term in Eq.~\eqref{eq:charge_eigenbasis}, and leads to phase errors. 

The longitudinal error term has a coupling strength of order $g_c$ modulated by the factor $d_{km}$, and contains frequencies $\omega \in \{\pm 2\Omega + (n \pm 1) \omega_d \}$, where $n \equiv  2k - m$ runs over all integers $\mathbb{Z}$. Spurious resonances occur when these frequencies vanish. The first of these happens at $n=0$, which leads to $2\Omega + \omega_d =0$, and consequently $\omega_r = 0$ when the charge qubit is driven at resonance (see Eqs.~\eqref{eq:charge_resonance}). This implies that a large resonator frequency helps preserve the validity of the RWA. The RWA condition $\abs{g_c d_{km}/\omega_r}\ll 1$ is further supported by the small magnitude of the modulating factor, $\abs{d_{km}} < 1$ when $n=0$. 

When $n \neq 0$, spurious resonances emerge at frequencies $2\Omega = \omega_r (1 \mp 1/n)$. Since $2\Omega \le \omega_r$, we ignore the expression with the positive sign and take $n \in \mathbb{Z}^+$. When $n$ increases, this physically implies that the charge qubit is driven at increasingly larger drive amplitudes, while the resonator-qubit detuning must decrease for the red-sideband resonance to remain valid. For progressively larger values of $n$, the modulating factor decreases. This can be observed by examining the behaviour of the binomial coefficients in $d_{km}$. When $m = 2k-n$ is substituted into the binomial coefficients, the ratio for successive values of $n$ increasing by 1 is given by
\begin{align}
	\frac{C^{4k}_{2k-(n+1)}} {C^{4k}_{2k-n} }= \frac{1}{(2k-(n+1))(2k+(n+1))}, \label{eq:binomial ratios}
\end{align}
which converges to zero as $n \rightarrow \infty$. In addition, the remaining term in the modulating factor, $\abs{\frac{C^{1/2}_{k}}{(2i)^{4k}}} \ll 1$ for all $k$, further reduces the overall contribution. As a result, this relaxes the strictness of the RWA condition when $n$ is large, since $\abs{\frac{g_c d_{km}}{2\Omega - \omega_r (1 \pm 1/n)}} \approx \abs{\frac{g_c d_{km}}{\Delta_c}} \ll 1$ is satisfied due to the modulating factor $d_{km}$ becoming vanishingly small as $n$ increases. In other words, even as resonator-qubit detuning approaches zero due to large values of $n$,  the RWA condition can still hold.

To summarize, in terms of the dimensionless frequencies, the RWA breaks down for small $n$ at
\begin{align}
    2 \tilde \Omega =  \tilde\omega_r\left(1 - \frac{1}{n}\right),  \label{eq:charge_qubit_RWA}
\end{align}
because the strictness of the RWA condition relaxed significantly when $n$ is large. In other words, for any value of resonator frequency, allowed values of drive amplitude fall within the intervals $2\tilde \Omega \in (0, \frac{\tilde\omega_r}{2}), (\frac{\tilde\omega_r}{2}, \frac{2 \tilde\omega_r}{3}), (\frac{2 \tilde\omega_r}{3}, \frac{3 \tilde\omega_r}{4}), \dots $, with successive intervals of decreasing range, up to the theoretical maximum of $2\tilde\Omega = \tilde \omega_r$. We shall see that this is indeed the case for the charge qubit in the next section where gate fidelities are calculated.

Finally, it is important to make a few comments on the frequencies of the longitudinal error term to address potential objections in advance. Firstly, it is not surprising that there are infinite frequency components, because the orbital detuning drive contains a square root which introduces  sharp cusps in the waveform. Although this results in an infinite number of spurious resonances with potentially significant consequences, we will show that the impact on gate fidelity is actually minimal in Sec.~\ref{sec:Gate time correction}. Secondly, an infinite set of frequencies implies very high bandwidth requirements in experimental settings. We address this in the Discussion, Sec.~\ref{sec:Bandwidth requirement}, demonstrating that state-of-the-art waveform generators can meet the requirements for driving the charge qubit.


\subsubsection{Fourier decomposition and RWA conditions of DQD spin qubit}\label{sec:RWA_spin_qubit}

We can analyze the Fourier components of the four spin qubit error terms in  Eqs.~\eqref{eq:spin_errors} in a similar way. Notably, three of the error terms -- $\hat V_\text{s,red}(t), \hat V_\text{s,blue}(t), \hat V_\text{s,long}(t)$ -- have analogous meanings to their counterparts in the charge qubit. The key difference is that the spin qubit has an additional error term, $\hat V_\text{s,drive}(t)$, which is absent in the charge qubit. This term is responsible for the most significant limitations on the drive parameters.

The first error term $\hat V_\text{s,red}^{(j)}(t)$ has coupling strength of order $g_s/2$ and contains frequencies $\omega \in \{\pm 2\omega_d \}$. Given the red-sideband resonance conditions in Eq.~\eqref{eq:spin_resonance}, these frequencies can also be expressed in terms of the spin qubit transition frequency, $\omega \in \{\pm 2\omega_s \}$. The second error term $\hat V_\text{s,blue}^{(j)}(t)$ has coupling strength of order $g_s$ and contains frequencies $\omega \in \{\pm 4\Omega, \pm(4\Omega + 2\omega_d) \} = \{\pm 2(\omega_r - \omega_s), \pm 2\omega_r ) \} $, where in the second equality we have used the red-sideband resonance conditions. The third error term $\hat V_\text{s,long}^{(j)}(t)$ has coupling strength of order $g_s$ and contains frequencies $\omega \in \{\pm 2\Omega, \pm(2\Omega + 2\omega_d) \} = \{\pm (\omega_r - \omega_s), \pm (\omega_r + \omega_s ) \} $. 

Taken together, for the RWA to hold, we need the conditions
\begin{gather}\label{eq:spin_RWA_lower}
	\abs{\frac{g_s}{4\omega_s}} = \abs{\frac{1}{4\tilde \omega_s}}  \ll 1,~
	\abs{\frac{g_s}{2\omega_r}} = \abs{\frac{1}{2\tilde \omega_r}} \ll 1, \nonumber \\
	\abs{\frac{g_s}{\omega_r+ \omega_s}} = \abs{\frac{1}{\tilde\omega_r+\tilde \omega_s}} \ll 1,~
	\abs{\frac{g_s}{\Delta_s}} = \abs{\frac{1}{\tilde\Delta_s}} \ll 1.
\end{gather}
These RWA conditions yield conclusions similar to that of the charge qubit. Specifically, the last condition implies that a large resonator-qubit detuning -- and therefore a large drive amplitude -- is beneficial, suggesting a lower limit on drive amplitude. In addition, because the resonator and spin qubit transition frequencies are in the microwave range, the RWA generally holds in practice -- unless ultrastrong coupling is achieved. These conclusions are similar to that of the charge qubit. However, current experiments with spin qubit-resonator devices have not reached the ultrastrong coupling regime, unlike the charge qubit case.

Finally, the $\hat V_\text{s,drive}(t)$ error term, which represent deviations from an ideal Rabi drive, is of magnitude $\Omega$ with Fourier components $\omega \in   \{ \pm 2\omega_d, \pm (2 \Omega + 2 \omega_d), \pm (2 \Omega - 2 \omega_d) \} = \{ \pm 2\omega_s, \pm ( \omega_r + \omega_s), \pm (  \omega_r -  3\omega_s) \} $. The additional RWA conditions in dimensionless frequencies, are 
\begin{align}\label{eq:RWA_spin_upper}
	\abs{\frac{\tilde\Omega} {\tilde\omega_r + \tilde\omega_s}} \ll 1, ~
	\abs{\frac{\tilde\Omega} {2\tilde\omega_s}} \ll 1, ~
	\abs{\frac{\tilde\Omega}{\tilde\omega_r -  3\tilde\omega_s}} \ll 1. 
\end{align}

These conditions all lead to an upper limit on  the drive amplitude. Since the RWA condition from the blue error term provides a lower limit to drive amplitude, when taken together, they imply that there could be an optimal drive amplitude between these bounds. In the next section, we compute gate fidelities numerically and see that this is indeed the case for the spin qubit.

\section{Numerical Results}\label{sec:Numerical Results}

\begin{figure*}
	\centering
	\includegraphics[width=\linewidth]{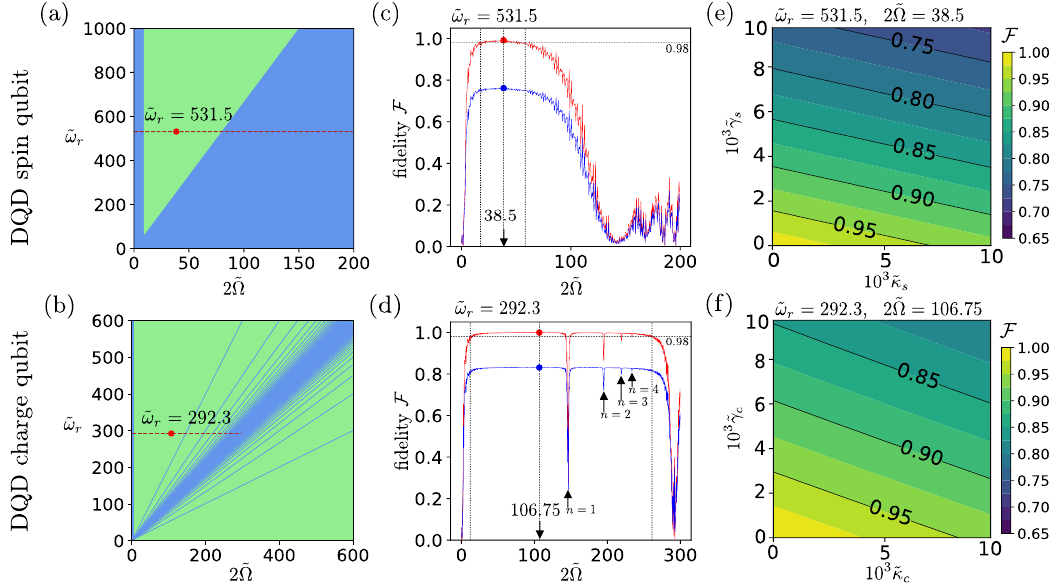}
	\caption{Gate fidelity of spin (top row) and charge (bottom row) qubits. (a, b) Map of high (green) and low (blue) fidelity regions against rescaled resonator frequency and drive amplitudes. Dashed red lines are linecuts taken and plotted in panels (c) and (d). (c, d) Fidelity linecuts at $\tilde\omega_r = 531.5$ and $292.3$, against rescaled drive amplitude, with (blue) and without (red) decoherence, for rates $\tilde\kappa_s = \tilde\kappa_c = \tilde\gamma_s = \tilde\gamma_c = 0.008$. Highest fidelities for spin and charge qubits are $99.20\%$ and $99.95\%$ (red circles in panels c, a), at optimal drive amplitudes $2\tilde\Omega = 38.5$ and $106.75$ respectively, indicated by the central dashed vertical line.
		With decoherence, optimal fidelities drop to  $76\%$ and $83\%$ (blue circles in panels d, b) for the spin and charge qubits. The boundaries of high-fidelity regions identified in panels (a) and (b) are shown as vertical dashed lines at the two sides computed from the infidelity approximation of Eq.~\eqref{eq:infidelity formula}. Since these boundaries intersect with the 98\% fidelity line, the analytical approximation shows excellent agreement with the numerical results. Arrows denoting $n=1 \dots 4$ indicate locations of spurious resonances from the multiple frequency components of the charge qubit detuning drive, as identified in Eq.~\eqref{eq:charge_qubit_RWA}.  The effect on fidelity is negligible beyond $n=4$. (e, f) Fidelity against rescaled photon loss and dephasing rates at optimal drive amplitudes and resonator frequencies used in panels (a--d), showing that the charge qubit is more resilient against noise than the spin qubit.}
	\label{fig:F2}
\end{figure*}


\subsection{Map of high-fidelity regions}\label{sec:Map of high-fidelity regions}

To map out high-fidelity regions in parameters space, we first define the fidelity between an ideal CZ gate and the actual gate for the charge/spin qubit as~\cite{Wang2008}
\begin{equation}
	\mathcal{F}_\text{c/s} = \frac{1}{d^2}\left| \text{Tr}\left( \mathscr{U}_\text{cz}^\dag \mathscr{U}_\text{c/s,exact} \right)\right|, \label{eq:fidelity}
\end{equation}
where $\mathscr{U}_\text{cz}$ is the superoperator corresponding to the ideal gate $\hat U_\text{cz}$,  and $\mathscr{U}_\text{c/s,exact}$ is the superoperator describing the exact evolution of the charge/spin qubit system, with $d=4$ being the dimension of the two-qubit system. We first compute the exact evolution (i.e. without RWA) to quantify the effect of the error terms on gate fidelity. We then incorporate decoherence in order to deduce the conditions for high-fidelity operation. The superoperator formalism and the decoherence model are detailed in the Methods section.

Earlier in Sec.~\ref{sec:RWA validity conditions and Fourier decomposition of error terms}, we established the RWA conditions that provided heuristic bounds for the frequencies of the system. Now, we rigorously refine these bounds with a Dyson series expansion of all the time-evolution operators, and compute the resulting infidelity in order to map out high-fidelity regions within the two-dimensional rescaled frequency space defined by $\{\tilde \omega_r, 2\tilde\Omega\}$. From the Dyson series expansions, we express $\mathcal{F}_\text{c/s} = |1-\delta_\text{c/s}|$, where gate infidelity is, in general, complex: $\delta_\text{c/s} \in \mathbbm{C}$. To leading order in the dimensionality of integration domain, gate infidelity is (Appendix~\ref{supp:perturbative expansions})
\begin{widetext}
\begin{align}
	\delta_\text{c/s}(\tilde \omega_r, \tilde\Omega) \approx& \delta_\text{c/s}^{(2)}(\tilde \omega_r, \tilde\Omega) \nonumber \\
	= &\frac{1}{4}  \Tr \prescript{}{r}{\bra{0}} \int_{\tau_\text{ini}(b)}^{\tau_\text{fin}(b)} d\tau_1 \int_{\tau_\text{ini}(b)}^{\tau_1} d\tau_2 \left[ \sum_{b<b^\prime} \hat V_\text{c/s}(b;\tau_1) \HR(b^\prime)  + \sum_{b>b^\prime} \left( \HR(b)\hat V_\text{c/s}(b^\prime;\tau_2)+ \hat V_\text{c/s}(b;\tau_1)\hat V_\text{c/s}(b^\prime;\tau_2)   \right) \right. \nonumber \\
	&  \left. \sum_b \left(\hat V_\text{c/s}(b;\tau_1) \HR(b) -\mathcal{T}\left( \{\HR(b), \hat V_\text{c/s}(b;\tau_2) \} \right) + \hat V_\text{c/s}(b;\tau_1) \hat V_\text{c/s}(b;\tau_2)  \right) \right]  \ket{0} _r, \label{eq:infidelity formula} 
\end{align}
\end{widetext}
where each error term in the formula is an independent sum of specific errors $\hat V_\text{c/s} \equiv \sum_k \hat V_{\text{c/s},k}$, with $k \in \{\text{red, blue, long, drive}\}$.

There are two key observations from this result. Firstly, all first order terms vanish for both the charge and spin qubit systems. The dominant infidelity contribution in the qubit systems thus scales with the second-order in the dimensionality of integration, ensuring that small errors remain negligible, a critical feature for scalable quantum gate performance. Secondly, the contributions to infidelity arise from cross terms between red-sidebands and error terms between stages and the effect of time ordering of non-commutative error terms.

Now, we estimate the boundaries of the high-fidelity regions by examining the behavior of the infidelity series expansion. While the Dyson series converges generally when the RWA is valid, the practical utility of the series in perturbation theory depends not just on convergence but on the monotonic decrease of its terms. Specifically, when the leading (second-order) term satisfies $|\delta_\text{c/s}^{(2)}(\tilde \omega_r, \tilde\Omega)| < \frac{1}{4}$, the sequence of higher-order terms tends to decrease monotonically, allowing the series to be truncated perturbatively with controlled error (Appendix~\ref{supp:perturbative expansions}). 
In other words, the preimage of this bound under the leading-order mapping identifies the parameter regions where the series behaves perturbatively and defines the high-fidelity domains.
As mentioned in the next section~\ref{sec:Gate fidelity results}, numerical results show that our imposed perturbative-infidelity condition ensures a fidelity higher than $98\%$. Henceforth, all mentions of "high-fidelity regimes" refer to a sub-parameter space with fidelity higher than $98\%$. The regimes can be further tightened by imposing a stricter bound on the R.H.S. of the condition.

Next, we summarize the results of the high-fidelity regions, the derivation of which are shown in Appendix~\ref{supp:perturbative expansions}. Firstly, as the rescaled drive amplitude $2 \tilde\Omega$ approaches zero for the spin qubit, the contribution to infidelity is dominated by error terms with frequencies in multiples of the drive amplitude. To leading order in $1/\tilde\Omega$, this yields $\delta_\text{s}^{(2)} \approx \frac{5}{8i\tilde\Omega} \Delta\tau_\text{total}$, where $\Delta \tau_\text{total} = (3+\sqrt{2})\pi$ is the total dimensionless gate time. We can then obtain the lower bound on drive amplitude for the spin qubit to be $|2\tilde\Omega| > \frac{5}{4} \Delta\tau_{\text{total}}$.

For the charge qubit, the blue-sideband error term contributes to infidelity in a similar way as the spin qubit's. The complication from the orbital detuning drive in the longitudinal term is considerably simplified because only the frequencies at $n \equiv 2k-m = \pm 1$ in Eq.~\eqref{eq:epsilon_expansion} dominate. Therefore, when drive amplitude $2\tilde\Omega$ approaches zero, the dominant infidelity contribution to leading order in $1/\tilde\Omega$ is $\delta_\text{c}^{(2)} \approx \frac{\Delta\tau_\text{total}}{i2\tilde\Omega}\left(\frac{1}{4} + 4 (2\xi_1)^2\right)$ where $\xi_1 \approx 0.196$ is the coefficient of the sum of the $n=\pm1$ exponentials in Eq.~\eqref{eq:epsilon_expansion}. We thus obtain the lower bound of the drive amplitude for the charge qubit as $ \abs{2\tilde\Omega} > 0.865 \Delta\tau_\text{total}$.

Next, we estimate the upper bounds on drive amplitude. For the spin qubit, we find that the drive error term dominates infidelity. Numerically, we find $2 \tilde\Omega  < \frac{2 \tilde\omega_r}{18.269}$, which upper bounds the drive amplitude of the spin qubit. For the charge qubit, as discussed in Sec.~\ref{sec:RWA_charge_qubit}, spurious resonances arise from two main terms. The first is the red-sideband error term when resonator-qubit detuning $\Delta_c \equiv \omega_r - \omega_c$ approaches zero as expressed in the RWA condition in Eq.~\eqref{eq:charge_RWA_upper}. 
The second is the longitudinal error term, at small integer multiples of the driving frequency parameterized by $n$ in Eq.~\eqref{eq:charge_qubit_RWA}. For the former, the red-sideband error term yields $\delta_\text{c}^{(2)} \approx \frac{9}{4i(\tilde\omega_r - 2\tilde\Omega)} \Delta \tau_\text{total}$, which establishes an upper bound for the drive amplitude $2\tilde \Omega < \tilde \omega_r - \tfrac{9}{4}\tau_\text{total}$. 
For the latter, Eq.~\eqref{eq:charge_qubit_RWA} explicitly identifies the center of the low-fidelity regions for small $n$; these correspond to narrow parameter regions. The width of these strips scale quadratically with the Fourier coefficient $\xi_n$ (see Appendix~\ref{supp:detuning drive}) given by Eq.~\eqref{eq:charge qubit widths} and become vanishingly small for higher order terms. 

To summarise, the bounds that map out the high-fidelity regions in the  two-dimensional parameter space  $(2\tilde\Omega, \tilde \omega_r)$ are 
\begin{widetext}
\begin{align}
	\frac{5}{4}  \Delta \tau_\text{total} <& 2\tilde \Omega <  \frac{2\tilde\omega_r}{18.269}  	 \text{   (spin qubit)},  \label{eq:bounds spin qubit} \\
	0.865 \Delta\tau_\text{total} < & 2\tilde \Omega < \tilde \omega_r - \frac{9}{4} \Delta \tau_\text{total}, \text{excluding neighborhoods of } 2\tilde \Omega= \tilde\omega_r \left(1 - \frac{1}{n}\right)	 \text{   (charge qubit)}. \label{eq:bounds charge qubit}
\end{align}
\end{widetext}

These boundaries are mapped out in Fig.~\ref{fig:F2}(a,~b) with green (blue) regions representing high (low) fidelity. The spin qubit has a single, triangular high-fidelity region with its width proportional to frequency. In contrast, the charge qubit has multiple high-fidelity regions which are symmetrical about the diagonal $2\tilde \Omega = \tilde \omega_r$. These high-fidelity slices become sharper and narrower around the diagonal, which appears to diminish the practicality of operating in those regions. Fortunately, as mentioned in the preceding discussion on the breakdown of RWA from the $\epsilon(t)$ drive, at the boundaries of these narrow slices, the loss in fidelity is very small. In addition, for the same $\tilde \omega_r$, the high-fidelity regions for the charge qubit are broader than that of the spin qubit, giving it greater flexibility in operation. In addition,  the plot is mirrored about the $2\tilde \Omega = \tilde\omega_r$ diagonal because the mirrored, lower portion represent negative detuned frequencies which are allowed  solutions for negative integer $n$ in Eq.~\eqref{eq:charge_qubit_RWA}.

\subsection{Gate fidelity results}
\label{sec:Gate fidelity results}

In this section, we compute the exact numerical calculation of gate fidelity through a linecut in phase space, as shown by the dashed red lines in Fig.~\ref{fig:F2}(a, b) and numerically verify that the high-fidelity boundaries from the analytical approximations in the preceding discussion are indeed accurate.

Taking experimental parameters from Ref.~\cite{Mi2018}, we compute gate fidelities using $\omega_r / 2\pi = 5.846 ~\mathrm{GHz}$, $g_s/2\pi = 11~\mathrm{MHz}$ and $g_c/2\pi =  40~\mathrm{MHz}$, corresponding to $\tilde \omega_r = 531.5$ for the spin qubit and  $\tilde \omega_r = 292.3$  for the charge qubit. Using these values, the partial set of RWA conditions that require resonator frequency to be much larger than coupling strength are satisfied. Qubit transition frequencies are constrained by drive parameters according to the resonance conditions; it is therefore important to study how fidelities change with drive parameters. For the charge qubit, the remaining RWA conditions require the rescaled drive amplitude to be much larger than unity, with the RWA violations occuring at specific ratios of the resonator frequency as described by Eq.~\eqref{eq:charge_qubit_RWA}, until the theoretical maximum of $\tilde\omega_r$. On the other hand, for the spin qubit, while the rescaled drive amplitude also has to be much larger than unity, it is upper bounded according to the last RWA condition in Eq.~\eqref{eq:RWA_spin_upper}. 

We calculate gate fidelities numerically and plot them in Fig.~\ref{fig:F2}(c,~d) against rescaled drive amplitude $2\tilde \Omega$. The numerical results verify the analyses of the exact Hamiltonians: both qubits have fidelities greater than $99\%$ when the RWA is satisfied. Comparing the two qubit systems, the charge qubit has wider optimal regions and flatter fidelity curves, demonstrating greater robustness across drive amplitudes. The drop in fidelity is greatest when $2\tilde \Omega = \tilde \omega_r$, and is successively smaller for larger values of $n$. Beyond $n=4$, the infidelity from spurious resonances at high frequencies of the detuning drive is negligible, confirming the analysis of Eq.~\eqref{eq:binomial ratios}. 
At optimal drive amplitudes, fidelities of $99.95\%$ and $99.20\%$ can be achieved for the charge and spin qubits.
 
We can draw two conclusions from this result. Firstly, the infidelity from error terms is much smaller for the charge qubit than the spin qubit, suggesting that reducing the number of error terms from parametric driving is potentially an important route towards better performance. Secondly, even without decoherence, error terms reduce fidelities to a level below the threshold for fault-tolerance, demonstrating the need to go beyond the usual RWA in the modeling of quantum dot-resonator systems.

With decoherence (see Appendix~\ref{sec:Methods}), the fidelity curves in Fig.~\ref{fig:F2}(c,~d) are similar, but optimal fidelities drop to $83\%$ and $76\%$ for the charge and  spin qubits, using dimensionless decoherence rates $\tilde\kappa_s = \tilde\kappa_c = \tilde\gamma_s = \tilde\gamma_c = 0.008$. These are much smaller than typical rates in quantum dot-resonator experiments, e.g. Ref.~\cite{Mi2018} reports $\gamma_c/2\pi = 35$~MHz, $\gamma_s = 5$~MHz and $\kappa / 2\pi = 1.3$~MHz, corresponding to  $\tilde \gamma_c = 1.750$, $\tilde \gamma_s = 0.455$, $\tilde \kappa_c = 0.065$ and $\tilde \kappa_s = 0.118$. However, with these decoherence rates, gate fidelities drop to below $50 \%$, suggesting that current decoherence rates and/or coupling strengths have to improve in order to realize viable quantum gates.

Next, at the optimal drive amplitudes for both qubit systems, we calculate gate fidelity against photon loss and qubit dephasing rates in Fig.~\ref{fig:F2}(e,~f). We observe that both qubits are more sensitive to dephasing than photon loss from the gradient of the contours. Extracting the gradient of the respective contour, we find that fidelities greater than $95\%$ require
\begin{align}
	\frac{\tilde\gamma_s}{1.554} + \frac{\tilde\kappa_s}{7.150} & \leq 10^{-3} ~\text{(spin qubit)}, \label{eq:ultrastrong coupling spin qubit}\\
	\quad \frac{\tilde\gamma_c}{2.927} + \frac{\tilde\kappa_c}{8.250} & \leq 10^{-3}~\text{(charge qubit)}. \label{eq:ultrastrong coupling charge qubit}
\end{align}
We take the $95\%$~\blue{\cite{tuckett2020fault}} threshold rather than the fault-tolerant benchmarks greater than $99.9\%$~\cite{Aliferis2007, Aliferis2008, Aharonov2008}, since current experimental devices have yet to come close to meeting these stringent requirements. For example, Refs.~\cite{Mi2017, Stockklauser2017, Mi2018, Samkharadze2018, Landig2018, vanWoerkom2018, Borjans2020, Harvey-Collard2022} reported rates  yielding figures  $1-3$ orders of magnitude above the threshold set by the inequalities. Notably, recent experiments showed improved rates, with the best rates in a DQD charge qubit-resonator device achieving $\sim 8\times 10^{-3}$ for the left-hand-side of inequality~\eqref{eq:ultrastrong coupling charge qubit} in Ref.~\cite{Scarlino2022}, resulting in a fidelity of $61\%$. Another recent experiment characterizing a high-impedance TiN resonator yielded a photon loss rate of $\kappa \approx 15\times 2\pi$~kHz~\cite{Holman2021}, which is an improvement by two orders of magnitude from earlier works. These results suggest an optimistic outlook for quantum dot-resonator systems. Another takeaway is that for the same decoherence rates, we observe that the charge qubit performs better than the spin qubit. This is due to the stronger charge-photon coupling and hence faster gates, as well as fewer error terms. 


\section{Discussion}\label{sec:Discussion}

Our analysis reveals that the charge qubit outperforms the spin qubit in CZ gate performance, achieving higher optimal fidelity, broader tolerance to drive parameter variations, and greater robustness under decoherence. To critically assess the validity of this conclusion, we address three key considerations, namely (i) bandwidth constraints of the charge qubit, (ii) further  optimization of the spin qubit, and (iii) alternative control schemes that may mitigate the limitations in either system.

\subsection{Bandwidth constraints of the charge qubit}\label{sec:Bandwidth requirement}

\begin{figure}
	\includegraphics[width=0.6\linewidth]{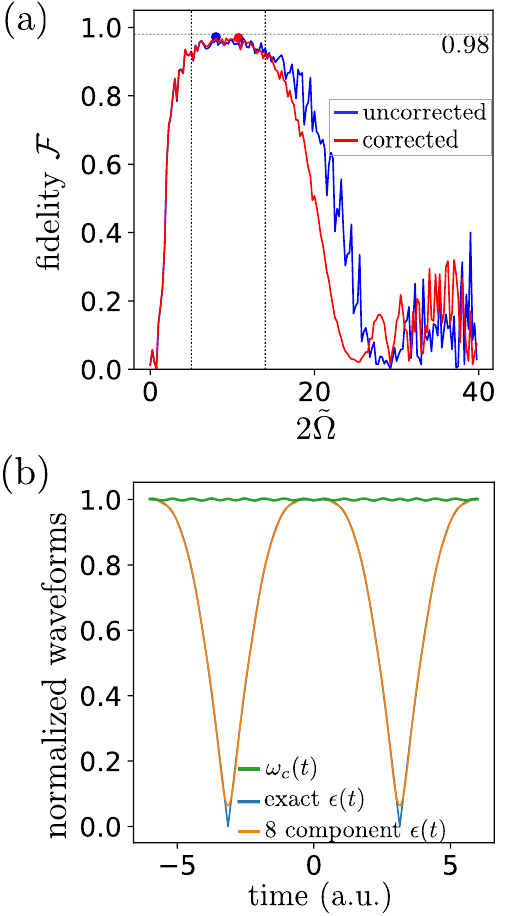}
	\caption{(a)  Spin qubit fidelity for $\tilde\omega_r = 158$ which is smaller than the linecut in Fig.~\ref{fig:F2}(c). Red (blue) curves represent fidelity with (without) gate time correction Appendix~\ref{supp:gate time correction}.  The curve is smoothed but the best fidelity (circles) is not improved.  (b)  Exact $\epsilon(t)$ waveform (blue) compared against realistic waveform (orange) composed with the first 8 Fourier components. The latter produces a good approximation of constant qubit frequency $\omega_c(t)$ (green).}
	\label{fig:F3}
\end{figure}

First, we evaluate whether the finite bandwidth of the charge qubit’s detuning drive could restrict achievable fidelity, potentially limiting its practical advantage. For example, a state-of-the-art waveform generator of, say 50~GHz bandwidth, can only compose 2 Fourier components of the detuning drive at the optimal drive amplitude shown in Fig.~\ref{fig:F2}(d). This will lead to large waveform deviations and lowered fidelity, but it can be mitigated by driving at a lower frequency. This requires a higher qubit frequency, which is dynamically tunable, as well as a larger drive amplitude for the same resonator frequency. Fortunately, the broad high-fidelity region of the charge qubit offers considerable flexibility in the choice of operating point. For instance, moving to a larger drive amplitude and qubit frequency $2\tilde\Omega = \tilde \omega_c =247$ gives a smaller drive frequency $\tilde \omega_d = 45.3$, which allows 8 Fourier components to be composed for the detuning drive. Including finite Fourier components leads to additional contributions on gate infidelity due to the oscillatory qubit spliting $\tilde\omega_c$. However, this contribution becomes negligible once a sufficiently large number $N$ of Fourier components is included, for two reasons.
\begin{enumerate}
	\item The oscillation amplitude decreases with increasing $N$. For $N=8$, the oscillatory amplitude $\sim 10^{-3} \cdot 2\tilde\Omega$, contributing to infidelity at the level of $\sim (10^{-3} \cdot 2\tilde\Omega)^2$. 
	\item The leading oscillatory frequency scales as $(N+1)\tilde\omega_d$, contributing $\sim ((N+1)\tilde\omega_d)^{-2}$ to the infidelity. 
\end{enumerate}
As such, $\omega_c$ can be well-approximated to be a constant as plotted in Fig.~\ref{fig:F3}(b). Therefore, the total contribution to infidelity of this term after including $N=8$ Fourier components $\sim 10^{-5}$ \%. A more complete discussion of the driving amplitude and infidelity are in Appendix~\ref{supp:detuning drive} and~\ref{supp:perturbative expansions}. 

Therefore, we find a reduction in fidelity to $99.29\%$ mostly by moving away from the optimal drive amplitude. This remains slightly higher than optimal spin qubit fidelity. Thus, the charge qubit is not limited by bandwidth constraints.

\subsection{Further  optimization of the spin qubit}\label{sec:Gate time correction}

Next, we investigate whether a conventional gate time correction strategy can enhance the spin qubit’s fidelity to bridge the performance gap. This is done by accounting for the Bloch-Siegert shift~\cite{Ashhab2007, Lu2012},  which is a shift in qubit transition frequency due to strong driving. Effectively, this is a gate time correction which accounts for and eliminates phase errors from the  drive error $\hat V_\text{s,drive}^{(j)}(t)$. Mathematically, this can be seen by moving into a time-dependent rotating frame defined by $\exp\left[-\tfrac{i\Omega}{2\omega_d}\sin(2\omega_d t + 2\phi)\hat\sigma_z \right]$, which has a convenient analytical form using the Jacobi-Anger identity $\exp\left[ i z \sin(\theta) \right] = \sum_{n=-\infty}^{\infty} J_n(z) e^{in \theta}$, where $J_n (z)$ are $n$-th Bessel functions of the first kind. In this frame, the coupling strength $g(t) = g_s \cos \left[ \frac{ \Omega }{ 2\omega_d } \sin(2\omega_d t + 2\phi) \right]$ is time-dependent and a gate time correction can be derived from requiring the same total rotation angle as before, $\int_{\tiz{\stage}'}^{\tiz{\stage}'} g(s) ds = g_s\Delta \tau_b$. Full mathematical details are shown in Appendix~\ref{supp:gate time correction}. Performing  this correction, we replot fidelity in Fig.~\ref{fig:F3}(a). Our analysis shows that while fast fidelity oscillations are suppressed, the overall fidelity remains unchanged, with the peak fidelity retaining its original value. This limitation arises because the method only compensates for one of the two time-dependent precession axes governed by $\hat V_\text{s,drive}(t)$ (see Sec.~\ref{sec:resonance spin qubit}). The unaccounted axis introduces nonlinear dynamics -- stemming from its coupling to the modulated drive parameters -- which persist as a dominant source of infidelity, preventing further enhancement of the gate performance with the present strategy.

\subsection{Alternative control scheme}\label{sec:Alternative control scheme}

\begin{figure}
	\includegraphics[width=0.6\linewidth]{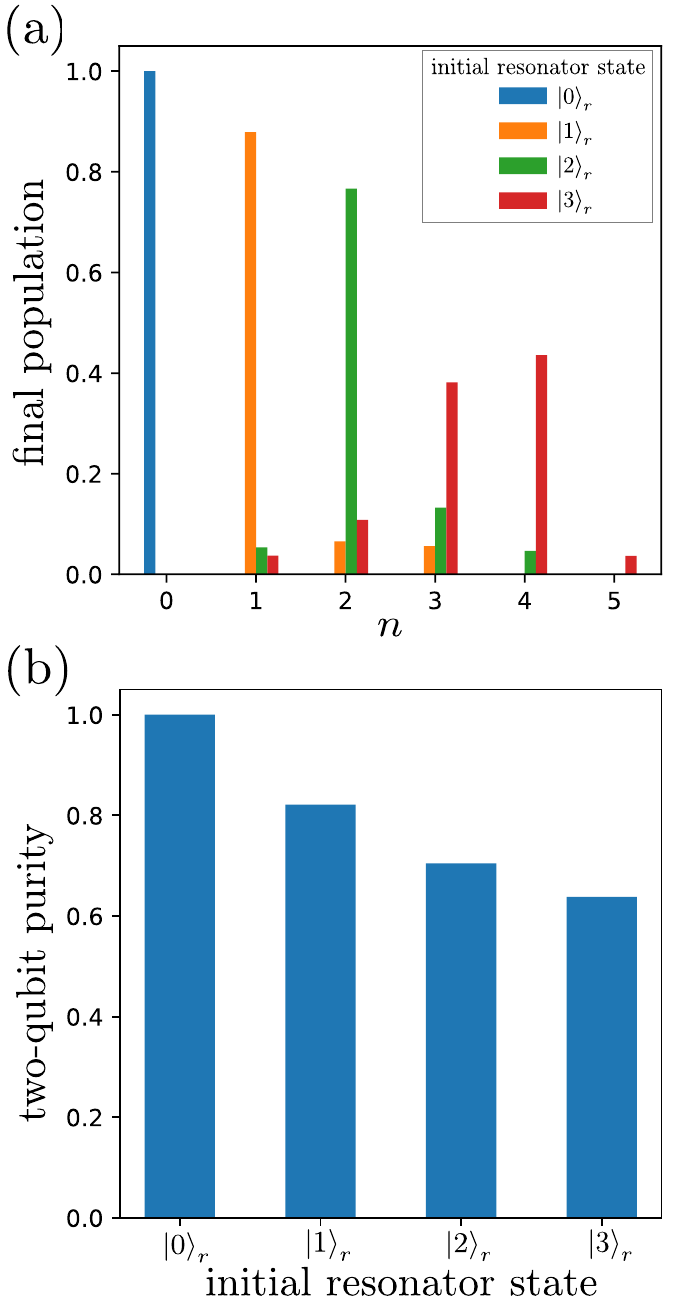}
	\caption{(a) Plot of final photon population of each Fock state $\ket{n}_r$ for various initial states. Only for an initial resonator vacuum state $\ket{0}_r$ does the gate sequence achieve a CZ gate and return to the vacuum subspace (see Appendix~\ref{supp:gate evolution}). (b) Plot of the purity of the final two-qubit state against initial resonator state. The two-qubit state is pure only with an initial vacuum state  $\ket{0}_r$, and becomes increasingly mixed as the initial resonator state contains a greater number of photons.} 
	\label{fig:F4}
\end{figure}


Finally, we investigate an alternative strategy -- driving the resonator rather than parametrically modulating the qubit, as described in Ref.~\cite{Srinivasa2016}. Our analysis reveals that the CZ gate protocol requires the qubits to initially reside in a state disentangled from the resonator’s vacuum (zero-photon) state. During the gate sequence, the protocol drives the system out of the zero-photon subspace, dynamically entangling the qubits with one- and two-photon resonator states, before eventually disentangling them to return to the target two-qubit subspace (see Appendix~\ref{supp:gate evolution} and Fig.~\ref{fig:F4}(a)). 
Crucially, any initial photon population disrupts this process by introducing non-CZ gate dynamics and produces residual entanglement between the qubits and the resonator.  As a result, the system collapses into a two-qubit mixed state upon projecting the resonator back to vacuum at the protocol’s conclusion. The purity of the two-qubit state is reduced with increasing initial photon population as shown in Fig.~\ref{fig:F4}(b). These effects result in gate infidelity that increases in relation to any initial photon population of non-zero resonator states. While resonator driving can generate sidebands analogous to parametric qubit modulation, it would also drive the resonator out of the zero-photon state, resulting in non-CZ gate dynamics and unavoidable loss in qubit purity. This renders resonator driving a suboptimal strategy for high-fidelity CZ gates.


\section{Conclusion}\label{sec:Conclusion}

In conclusion, we investigated the CZ gate for DQD spin and charge qubits, mediated by photons in a microwave resonator, using an existing gate sequence proposed in Ref.~\cite{Abadillo-Uriel2021}, with an additional zero-photon projection at the conclusion of the sequence. To achieve the gate, each qubit system was parametrically driven to achieve a red-sideband transition, in a working regime where the qubit transition frequency was detuned from the resonator frequency. We proposed a novel parametric drive for the charge qubit and compared it with an existing protocol for the spin qubit~\cite{Srinivasa2016, Srinivasa2024}. By analyzing the exact Hamiltonians that were induced, without the conventional RWA, we showed that our proposed drive on the charge qubit gave rise to fewer undesired time-dependent terms compared to the conventional drive on the spin qubit. Although this came at a cost of higher bandwidth requirement, we showed that bandwidth was not a limiting constraint. 

More significantly, our analysis of gate fidelity for both qubit systems revealed a non-trivial dependence on drive parameters that is absent under the RWA, where fidelities would appear ideal and independent of drive parameters.  Our analyses also showed that implementation details in parametric driving are critical: with the drive protocol that we proposed, the charge qubit had one fewer error term, resulting in a much smaller fidelity loss of 0.05\%, compared to 0.80\% for the spin qubit. We mapped out the subspace (non-trivial) in dimensionless parameter space where reasonably high-fidelity gate operations could be implemented by considering the exact evolution.  This was done in the space of independent system and drive frequencies rescaled by coupling strengths, to achieve overall generality. When the qubit systems were compared, we found that the charge qubit had superior optimal gate fidelities and a wider operating range that was more tolerant to variations in the drive parameter than the spin qubit. In addition, the charge qubit was more robust against qubit dephasing and photon loss. However, using experimentally feasible coupling strengths decoherence rates, we find that achieving a 95\% gate fidelity would be challenging for current devices.

One can distinguish between two primary contributions to infidelity: those arising from counter-rotating terms, and those resulting from decoherence. The effects of counter-rotating terms manifest as unitary errors, which, in practice, can be treated as calibration errors. Importantly, these errors are, in principle, correctable using more sophisticated techniques specifically designed to mitigate counter-rotating effects. Such approaches include pulse shaping~\cite{Buterakos2021, Barnes2022}, commensurate pulse calibration~\cite{Rower2024}, or hardware-level improvements~\cite{Rower2024}. While the latter methods are actively explored in superconducting circuit platforms, to our knowledge they have not yet been applied to hybrid semiconductor qubit systems. Extending these techniques to semiconductor qubits represents a promising avenue for further enhancing gate performance.

Decoherence, on the other hand, calls for device designs which are inherently quieter and cleaner materials growth, although it can also be mitigated with dynamical gate corrections~\cite{Buterakos2021, Barnes2022}. Strategies to reduce decoherence rates and/or improve coupling strengths must be pursued to push these limits further. Encouragingly, recent experiments demonstrating a loss-loss resonator~\cite{Holman2021},  ultrastrong coupling of a DQD charge qubit~\cite{Scarlino2022} indicate that these challenges may be overcome in practice. Whether operating in the strong or ultrastrong coupling regimes, our results highlight the necessity of going beyond the RWA to accurately understand and optimize long-range gate operations in DQD qubits. Together, these insights provide both a practical and conceptual roadmap for achieving higher-fidelity gates in emerging semiconductor qubit architectures.

\section{Acknowledgements}

This research was supported by the Ministry of Education, Singapore, under its Academic Research Fund Programme (MOE-T2EP50222-0017 and RG154/24) and the National Research Foundation, Singapore and A*STAR under its Quantum Engineering Programme 2.0 (NRF2021-QEP2-02-P07).

\begin{widetext}
\appendix

\section{Methods}\label{sec:Methods}


\subsection{Superoperator Formalism}\label{sec:Superoperator formalism}


In general, a superoperator $\mathscr O$ can be represented by a matrix with $(\mu, \nu)$ elements
\begin{align}\label{eq:superoperator elements}
	\left(\mathscr{O}\right)_{\mu\nu}  = \text{Tr} \left( \opvec_\mu^\dagger \mathscr{O} [\opvec_\nu] \right),
\end{align}
where $\{\opvec_\nu \}$ is an orthonormal basis of operators in the space of linear operators acting on a $d$-dimensional Hilbert space. The normalization condition is $\text{Tr}\left[ \opvec_\mu^\dagger \opvec_\nu \right] = \delta_{\mu\nu}$. On the right-hand-side, $\mathscr{O} [\opvec_\nu]$ denotes the superoperator $\mathscr{O}$ acting on the operator $\opvec_\nu$. The superoperator in Liouville space has dimensions $d^2 \times d^2$, reflecting the $d^2$ elements of the operator basis in a $d$-dimensional Hilbert space. For clarity, script font is used exclusively to represent superoperators. 

In the context of our system, we now define the operator basis $\{\opvec_\nu \}$. We denote by $\opvec^{(q)} = \frac{1}{\sqrt{2}}\begin{pmatrix} \I^{(q)}, & \X^{(q)}, & \Y^{(q)}, & \Z^{(q)}\end{pmatrix}$ the vector collecting the four rescaled Pauli operators of qubit $q$ where $q \in \{1,2\}$. Here, $\I^{(q)}$ is the identity operator in the qubit Hilbert space and rescaling normalizes the basis operators. Next, we denote the basis operators for photons as $\opvec^{(p)}_{mn} = \ket{m}_r \bra{n}, \text{ for } m, n \in \{0, \dots, N\}$ where $\ket{m}_r$ are photon number states in the resonator. In principle, the Fock space is infinite-dimensional but for computational purposes, we truncate at finite $N$. 

We use $\opvec^{(p)}$ and $\opvec^{(q)}$ to express the Hamiltonian of the full system. To this end, it is useful to introduce a single vectorization index $\lambda$ which will allow us to keep track of products of operators on the qubits with operators on the photon.
For any $({mn}, \mu_{1}, \mu_{2})$ which label the operator of the photon and the qubits respectively, we define a composite index $\lambda$ in a one-to-one correspondence such that  $\opvec_{\lambda} = \opvec^{(p)}_{mn} \opvec^{(1)}_{\mu_1} \opvec^{(2)}_{\mu_2}$.

Any operator $\hat O$ of interest for our system can be expressed as $\hat O = \sum_\lambda O^\lambda \opvec_{\lambda}$, and we can express the red-sideband and exact Hamiltonians, vectorized by $\vec{h}_\text{R}$ and $\vec{h}_\text{c/s,exact}$, as
\begin{align} 
	\HR (b) &= \sum_\lambda h_\text{R}^\lambda(b) \, \opvec_\lambda, \\
	\hat H_\text{c/s,exact}(b;t) &= \sum_\lambda h_\text{c/s,exact}^\lambda(b;t) \, \opvec_\lambda. \label{eq:exact H}
\end{align}

\subsubsection{Ideal CZ superoperator}

The action of an ideal CZ gate superoperator on a density matrix $\hat\rho$ is given by $\mathscr{U}_\text{cz} \left[ \hat\rho \right]= \hat U_\text{cz} \hat\rho \hat U_\text{cz}^\dag$. In the context of the gate sequence in Eq.~\eqref{eq:Ucz}, the ideal CZ gate superoperator can be expressed as the time-ordered product of superoperators,
\begin{align} \label{eq:ideal CZ superoperator}
	\mathscr{U}_\text{cz} =& \mathscr{M}_0 \mathscr{Z} \left( \mathcal T_{\leftarrow}\prod_{b=1}^{5} \mathscr{S}_\text{red}(b) \right).
\end{align}
where $\mathcal T_{\leftarrow}$ performs the normal time-ordering for the five qubit-photon stages, $b = 1\dots 5$. Here, $\mathscr{S}_\text{red}(b)$ denotes the superoperator describing the ideal red-sideband evolution of stage $b$ of the charge/spin qubit system, $\mathscr{Z}$ is the single-qubit superoperator and $\mathscr{M}_0$ is the zero-photon projection superoperator. These are
\begin{align}
	&\mathscr M_0 [\hat\rho] = \hat M_0 \hat\rho \hat M_0 \label{eq:projection superoperator}, \\
	&\mathscr Z [\hat\rho] = \hat Z^{(1)} (\tfrac{\pi}{\sqrt{2}}) \hat Z^{(2)} (-\tfrac{\pi}{\sqrt{2}}) \hat\rho \hat Z^{(2)} (-\tfrac{\pi}{\sqrt{2}})^\dag \hat Z^{(1)} (\tfrac{\pi}{\sqrt{2}})^\dag \label{eq:Z superoperator}, \\
	&\mathscr{S}_\text{red}(b) [\hat\rho] = \SR(b) \hat\rho \, \SR(b)^\dag.
\end{align}

\subsubsection{Exact CZ superoperator with and without decoherence}

The exact time-evolution superoperator for the charge/spin qubit is
\begin{align}
	\mathscr{L}_\text{c/s} \left[ \hat \rho\right] \equiv  \left( \mathscr{L}_{\text{c/s},U} +  \mathscr{L}_{\text{c/s},D} \right) \left[ \hat \rho\right] = \frac{d\hat\rho}{dt},
\end{align}
generates unitary time-evolution through the first term on the right side of the first equality
\begin{align}
	\mathscr{L}_{\text{c/s},U} \left[ \hat \rho\right] &\equiv -i [ \hat H_\text{c/s,exact}, \hat \rho ] 
\end{align}
and dissipative semigroup evolution through the second term
\begin{align}\label{eq:lindblad generator}
	\mathscr{L}_{\text{c/s},D} \left[ \hat \rho\right] &\equiv \sum_{ \ell} \gamma_{\text{c/s},\ell} \left(\hat{L}_{\text{c/s},\ell} \hat \rho  \hat{L}_{\text{c/s},\ell}^\dag - \tfrac{1}{2} \{\hat{L}_{\text{c/s},\ell}^\dag \hat{L}_{\text{c/s},\ell}, \hat \rho \} \right)
\end{align}
via the dissipative operators $\hat L$ and rates $\gamma$. Setting all the dissipative rates to zero $\gamma = 0$, i.e. $\mathscr{L}_D = 0$, allows us to compute the exact time-evolution without decoherence. When the rates are non-zero, we can compute the semigroup evolution with decoherence.  

The formal solution to a time-independent generator $\mathscr{L}$ is $\hat \rho(t) = e^{\mathscr{L}  t} [\hat\rho(0)]$.  Generalizing this to the time-dependent case allows us to find the exact time-dependent superoperator $\mathscr{S}_\text{c/s,exact}(b)$ of each $b$-th stage in the CZ gate sequence. During the $b$-th stage, the density matrix evolves from time $t_\text{ini}(b)$ to $t_\text{fin}(b)$ under a time-dependent generator $\mathscr{L}(b; t)$ as 
\begin{align} 
	\hat \rho (t_\text{fin}(b)) &= \mathscr{S}_\text{c/s,exact}(b) \left[\hat \rho (t_\text{ini}(b)) \right],\\
	\text{where }   \mathscr{S}_\text{c/s,exact}(b) &= \mathcal{T}_{\leftarrow} \exp \left[ \int_{t_\text{ini}(b)} ^{t_\text{fin}(b)} dt \mathscr{L}_\text{c/s}(b; t) \right] \label{eq:exact gate superoperator} \\
	&\approx \mathcal{T}_{\leftarrow} \prod_{j=1}^{k} \exp\left[ \mathscr{L}_\text{c/s} (b; t_j) \delta t  \right]. \label{eq:superoperator approximation}
\end{align}
In the final line, we have discretized the time steps with $\delta t \equiv \frac{t_\text{fin}(b) - t_\text{ini}(b)}{k-1}$. The approximation becomes exact when $k \rightarrow \infty$. We use Eq.~\eqref{eq:superoperator approximation} for numerical calculations in this paper.

\subsubsection{Modeling decoherence}\label{sec:modeling decoherence}

Decoherence is modelled by the dissipative operators, which we describe  in the Schr\"{o}dinger picture here. We take into account dephasing of qubits 1 and 2 as well as photon loss in the resonator. These are given by the Lindblad jump operators $\Z^{(1)}$ and $\Z^{(2)}$ for the charge qubit system, and correspondingly, $\tilde\sigma_z^{(1)}$ and $\tilde\sigma_z^{(2)}$ (with the tilde symbol) for the spin qubit system. The Lindblad jump operator for photon loss is identical for both spin and charge qubits and is given by $\a$, with photon loss rate $\kappa$. We assume dephasing rates for qubits 1 and 2 to be identical for each qubit system: $\gamma_\text{c}$ for the charge qubit and $\gamma_\text{s}$ for the spin qubit. Because single-qubit rotations are very fast in typical quantum dot systems (sub-ns) compared to qubit-resonator gate times (sub-$\mathrm{\mu}$s). Therefore, we assume all decoherence effects arise during the qubit-photon interaction stages.

\subsubsection{Lindblad operators in rotating frames}

Transforming into the rotating frames of the spin and charge qubits as described by the exact Hamiltonians Eqs.~\eqref{eq:exact H_charge},~\eqref{eq:exact H_spin}, the dissipative operators in the preceding section which were given in the Schr\"{o}dinger picture, gain time dependences. Using indices $j$ and $j^\prime$ to indicate coupled and uncoupled qubits and $r$ for the resonator, at stage $b$, the Lindblad operators in the rotating frames for the charge qubit are
\begin{align}
	\hat L_{\text{c},j} = \Z^{(j_b)}, ~
	\hat L_{\text{c},j^\prime} = \Z^{(j_b^\prime)}, ~
	\hat L_{\text{c},r} = e^{-i \omega_r t} \a.
\end{align}
For the spin qubit system, they are
\begin{gather}
	\hat L_{\text{s},j} =  - \sin(2 \Omega t) \tilde\sigma_x^{(j_b)} - \cos(2 \Omega t) \tilde\sigma_y^{(j_b)}, \nonumber \\
	\hat L_{\text{s},j^\prime} = -\tilde\sigma_y^{(j_b^\prime)}, ~
	\hat L_{\text{s},r} = e^{-i \omega_r t} \a. 
\end{gather}
The time-dependence in the photon loss rate arises from the frame rotating with the resonator frequency. The dephasing operators for both the coupled and uncoupled spin qubits are rotated by the final rotating frame of Eq.~\eqref{eq:R_{4,s}}. Additionally, the dephasing operator for the coupled spin qubit acquires a time-dependence from the additional time-dependent rotating frame of Eqs.~\eqref{eq:R_{3,s}}.

\subsection{Numerical convergence}

To ensure convergence of the numerical evaluation of Eq.~\eqref{eq:superoperator approximation}, the total number of time steps assigned is  proportional to the maximum oscillation frequency in the evolution. We assign 64 time instances for the shortest period of oscillation so that for each stage $b$, the total number of discretized time instances is $k = \left\lceil \frac{\max(\Delta \tau_b) \tilde \omega_{\text{max}}}{2 \pi} \times 64 + 1 \right\rceil$ where the ceiling operation $\lceil ~\rceil$ rounds up to the nearest integer. Using 64 samples for the shortest period, fidelity calculations converge to a precision of $\sim 5 \times 10^{-6}$. In the paper, we report fidelities to a precision of $10^{-4}$, well within convergence limits. For the charge qubit, there are high frequency terms arising from the detuning drive, but we found that frequencies higher than maximum of the spin qubit's could be neglected since their amplitudes are vanishingly small. Therefore, we could take the same sampling rate as that for the spin qubit while retaining numerical convergence.

\section{Ideal CZ gate evolution}\label{supp:gate evolution} 

Here, we detail the qubit-photon evolution stages corresponding to the CZ gate sequence of Eq.~\eqref{eq:Ucz} in the main text. 
The unitary time-evolution propagator corresponding to the red-sideband Hamiltonian of main text Eq.~\eqref{eq:red-sideband Hamiltonian} in the subspace of the $j$-th qubit and the photon Fock space can be expressed with the (coupled) qubit and photon operators as

\begin{subequations}
		\begin{align}
			& \SR^{(j)}(g \Delta t, \phi) = \exp \left [ -i\Delta t \HR^{(j)} (g,\phi) \right ] \nonumber \\
			&= \cos\left( \tfrac{ g \Delta t}{2} \sqrt{\a\c}\right) \ket{1}_j\bra{1} +  \cos\left( \tfrac{ g \Delta t}{2} \sqrt{\c\a}\right) \ket{0}_j\bra{0}   
			- \a \frac{\sin\left(\tfrac{g \Delta t}{2} \sqrt{\c\a} \right)}{\sqrt{\c\a}} e^{i\phi} \ket{1}_j\bra{0}
			+ \c \frac{\sin\left(\tfrac{g \Delta t}{2} \sqrt{\a\c} \right)}{\sqrt{\a\c}} e^{-i\phi} \ket{0}_j\bra{1} \\
			&= \cos\left( \tfrac{ g \Delta t}{2} \sqrt{1+ \hat n }\right) \ket{1}_j\bra{1} +  \cos\left( \tfrac{ g \Delta t}{2} \sqrt{\hat n}\right) \ket{0}_j\bra{0}   
			- \a \frac{\sin\left(\tfrac{g \Delta t}{2} \sqrt{\hat n} \right)}{\sqrt{\hat n}} e^{i\phi} \ket{1}_j\bra{0}
			+ \c \frac{\sin\left(\tfrac{g \Delta t}{2} \sqrt{1+\hat n} \right)}{\sqrt{1+\hat n}} e^{-i\phi} \ket{0}_j\bra{1} \label{eq:S_red}
		\end{align}
\end{subequations}
where $\hat n = \c\a$. 
Starting with a product state comprising a general two-qubit state $\ket{\psi} = c_{00} \ket{00} + c_{01} \ket{01} + c_{10} \ket{10} + c_{11} \ket{11}$ and the resonator vacuum state $\ket{0}_r$, the evolution stages are as follows.

\begin{subequations}
	\begin{align}
		\ket{\psi} \otimes \ket{0}_r &=  \left( c_{00} \ket{00} + c_{01} \ket{01} + c_{10} \ket{10} + c_{11} \ket{11} \right) \otimes \ket{0}_r  \nonumber \\
		& \xrightarrow{\SR(1)} \left( c_{00} \ket{00} + c_{10} \ket{10} \right) \otimes \ket{0}_r +  \left( c_{01} \ket{00} + c_{11} \ket{10}  \right) \otimes \ket{1}_r  \\
		& \xrightarrow{\SR(2)}  \left( c_{00} \ket{00} + \frac{c_{01}+c_{10}}{\sqrt{2}} \ket{10} \right) \otimes \ket{0}_r +  \left( \frac{c_{01}-c_{10}}{\sqrt{2}}  \ket{00} + c_{11} \cos\left(\frac{\pi}{2\sqrt{2}}\right) \ket{10}  \right) \otimes \ket{1}_r \nonumber \\
		& ~~~~~~~~~~~ - c_{11} \sin\left(\frac{\pi}{2\sqrt{2}}\right)\ket{00}  \otimes \ket{2}_r  \\
		& \xrightarrow{\SR(3)} \left[ c_{00}\ket{00} + \left( \frac{(c_{01}+c_{10}) \cos\left(\tfrac{\pi}{\sqrt{2}}\right) -i (c_{01}-c_{10}) \sin\left(\tfrac{\pi}{\sqrt{2}}\right)}{\sqrt{2}} \right) \ket{10} \right] \otimes \ket{0}_r \nonumber \\  
		& ~~~~~~~~~~~			+	\left[ \left( \frac{(c_{01}-c_{10}) \cos\left(\tfrac{\pi}{\sqrt{2}}\right) -i (c_{01}+c_{10}) \sin\left(\tfrac{\pi}{\sqrt{2}}\right)}{\sqrt{2}} \right)  \ket{00} - c_{11} \cos\left(\frac{\pi}{2\sqrt{2}} \right) \ket{10} \right] \otimes \ket{1}_r \nonumber \\
		& ~~~~~~~~~~~			+	 c_{11} \sin\left( \frac{\pi}{2\sqrt{2}} \right) \ket{00} \otimes \ket{2}_r \\
		& \xrightarrow{\SR(4)} \left( c_{00} \ket{00} + c_{10}e^{-\tfrac{i \pi}{\sqrt{2}}}  \ket{10} \right) \otimes \ket{0}_r +  \left( c_{01} e^{-\tfrac{i \pi}{\sqrt{2}}} \ket{00} - c_{11} \ket{10}  \right) \otimes \ket{1}_r  \\
		& \xrightarrow{\SR(5)} \left( c_{00} \ket{00} + c_{01}e^{-\tfrac{i \pi}{\sqrt{2}}} \ket{01} + c_{10}e^{\tfrac{i \pi}{\sqrt{2}}} \ket{10} - c_{11} \ket{11} \right) \otimes \ket{0}_r \\
		&  \xrightarrow{Z^{(1)}\left(\tfrac{\pi}{\sqrt{2}}\right) Z^{(2)}\left(-\tfrac{\pi}{\sqrt{2}}\right)} \left( c_{00} \ket{00} + c_{01}\ket{01} + c_{10} \ket{10} - c_{11} \ket{11} \right) \otimes \ket{0}_r.
		\end{align}
\end{subequations}

We note that the initial resonator state has to be a vacuum state -- any other state will not produce a CZ gate. 
Therefore, if the initial resonator state is a superposition of photon number states, a final projection onto the vacuum state is necessary to achieve a CZ gate.


\section{Analytical approximation of detuning drive}\label{supp:detuning drive}

The detuning drive on the DQD charge qubit is a square-root oscillatory function which was decomposed in Eq.~\eqref{eq:epsilon_expansion} of the main text. Here, we show that it has an alternative analytical approximation given by a Fourier cosine series~\cite{magnus1966formulas}
\begin{align}
	\epsilon(t) \approx \xi_{0} + \sum_{n=1}^{N} 2 \xi_{n} \cos\left(n(\omega_d t + \phi) \right) \label{eq:epsilon fourier cosine}.
\end{align}
for sufficiently large integer $N$. 
The approximation is exact in the limit $N \rightarrow \infty$. 
The amplitude of each frequency is given by the formula
\begin{align}
	\xi_n &= \frac{1}{2\sqrt{2} \pi} (-1)^{4c_n \mp n + 1} \frac{\Gamma\left(c_n - \frac{1}{2}\right) \Gamma\left(c_n + \frac{1}{4}\right) \Gamma\left(c_n + \frac{1}{2}\right) \Gamma\left(c_n + \frac{3}{4}\right)}{\Gamma\left(c_n - \frac{n}{2} + \frac{1}{2}\right) \Gamma\left(c_n - \frac{n}{2} + 1\right) \Gamma\left(c_n + \frac{n}{2} + \frac{1}{2}\right) \Gamma\left(c_n + \frac{n}{2} + 1\right)}  \nonumber \\
	& \times {}_5 F_4\left(\begin{matrix}c_n - \frac{1}{2} &c_n + \frac{1}{4} &c_n + \frac{1}{2} &c_n + \frac{3}{4} &1 \\ c_n - \frac{n}{2} + \frac{1}{2}& c_n - \frac{n}{2} + 1 &c_n + \frac{n}{2} + \frac{1}{2} &c_n + \frac{n}{2} + 1\end{matrix}; 1\right) \label{eq: sqrt wave coefficient}\; 
\end{align}
where $c_n = \lceil{n/2}\rceil$ and 
\begin{align}
	\prescript{}{p}{F}_q \left(\begin{matrix} a_1  &a_2  &\dots &a_p \\ b_1 & b_2  & \dots &b_q  \end{matrix}; z\right)  := \sum_{n=0}^{\infty} \frac{(a_1)_n \dots (a_p)_n}{(b_1)_n \dots (b_q)_n} \frac{z^n}{n!}
\end{align}
is the generalized hypergeometric function, and $(a)_n$ are Pochhammer symbols.  An explicit calculation of the first two amplitudes yield 
\begin{align}
	\xi_0 = 1 \times {}_5 F_4\left(\begin{matrix}-\frac{1}{2} &\frac{1}{4} &\frac{1}{2} &\frac{3}{4} &1 \\ 1& \frac{1}{2} &1 &\frac{1}{2}\end{matrix}; 1\right) \approx 0.731, ~~ \;
	\xi_1 = \frac{1}{8} {}_5 F_4\left(\begin{matrix}\frac{1}{2} &\frac{5}{4} &\frac{3}{2} &\frac{7}{4} &1 \\ 1& \frac{3}{2} &2 &\frac{5}{2}\end{matrix}; 1\right) \approx 0.196.
\end{align}


\section{Conventional gate time correction for spin qubit}\label{supp:gate time correction}

Here, we describe the mathematical details of the conventional gate time correction arising from the Bloch-Siegert shift~\cite{Ashhab2007, Lu2012}, which is a shift in qubit transition frequency due to strong driving. The correction eliminates phase errors from the strong drive by eliminating the $\Zj$ terms of the drive error term $\hat V_\text{s,drive}^{(j)}(t)$ in Eq.~(41--42) of the main text. Effectively, the phase evolution caused by the drive error term is  accounted for with the corrected gate time.

This is performed by first moving into a time-dependent rotating frame~\cite{Ashhab2007, Lu2012},
\begin{align}
	\hat R_{5,\text{s}}(t) = \exp[-\frac{i\Omega}{2\omega}\sin(2\omega t + 2\phi)\Zj].
\end{align}
Using the Anger-Jacobi identity,
\begin{align}
	&\exp\left[ i z \sin(\theta) \right] = \sum_{n=-\infty}^{\infty} J_n(z) e^{in \theta},
\end{align}
where $J_n(z) = \frac{1}{2\pi} \int_{-\pi}^{\pi}dx \cos(nx - z\sin x)$ are Bessel functions of the first kind, 
the exact spin qubit Hamiltonian in this fifth and final rotating frame can be evaluated, yielding
\begin{align}
	& \hat H_\text{s,rf5}^{(j)}(t)  \nonumber \\
	&=   \Omega e^{2i\Omega t + i z \sin(2\omega_d t + 2\phi)} \sin(2\omega_d t + 2\phi) \pj + \Omega e^{-2i\Omega t - i z \sin(2\omega_d t + 2\phi)} \sin(2\omega_d t + 2\phi) \mj \nonumber \\
	&-g \a \Bigg( e^{-i\omega_d t + i z \sin(2\omega_d t + 2\phi)} \sin(\omega_d t + \phi) \pj + e^{-i(4\Omega + \omega_d) t - i z \sin(2\omega_d t + 2\phi)} \sin(\omega_d t + \phi) \mj + e^{-i(2\Omega + \omega_d) t} \cos(\omega_d t + \phi)\Zj\Bigg) \nonumber \\
	&-g \c \Bigg( e^{i(4\Omega + \omega_d) t + i z \sin(2\omega_d t + 2\phi)} \sin(\omega_d t + \phi) \pj + e^{i\omega_d t - i z \sin(2\omega_d t + 2\phi)} \sin(\omega_d t + \phi) \mj + e^{i(2\Omega + \omega_d) t} \cos(\omega_d t + \phi) \Zj\Bigg),
\end{align}
where $z = \Omega / 2\omega_d > 0$. Factorizing the common function $g(t) = g \cos \left[ z \sin(2\omega_d t + 2\phi) \right]$ from sideband transition terms $\a\pj$ and $\c\mj$, and assuming the RWA by dropping the rest of the time-dependent terms, we obtain the red-sideband Hamiltonian with a time-dependent effective coupling $g(t)$,
\begin{align}
	\hat H_\text{s,rf5}^{(j)}(t) \approxRWA  ~ \HR^{(j)}(g(t), \phi) = \frac{ig(t)}{2} (e^{i \phi}\a\pj - e^{- i \phi} \c\mj).
\end{align}
To achieve the CZ gate, we find the corrected gate times of each stage with 
\begin{align}
	\int_{\tiz{b}'}^{\tfz{b}'} ds \cos\left[ z \sin(2\omega_d s + 2\phi_b) \right] = \Delta \tau_b.
\end{align}
We can find the corrected gate time $\Delta \tau'_b = {\tfz'} - {\tiz'}$ for each $b$-th stage that is related to the original gate times $\Delta \tau_b$ by evaluating the above integral, which yields
\begin{align}
	J_0 (z) \Delta \tau'_b + \sum_{n = 1}^{\infty} \frac{J_{2n}(z)}{2n \omega_d } \sin \left[ 2n(2\omega_d t + 2\phi_b) \right] \Big|^{\tiz{b}' + \Delta \tau'}_{\tiz{b}'} = \Delta \tau_b.
\end{align}
The integral is evaluated by the Fourier cosine series of the real function $f(x)=\cos( z \sin x ), z \in \mathbbm{R}$, i.e.,
\begin{align}
	\cos( z \sin x ) = J_0(z) + 2 \sum_{n=1}^{\infty} J_{2n}(z) \cos(2nx)
\end{align}

However, numerical results shown by Fig.~3(a) of the main text show no improvement in fidelity, due to the remaining qubit raising and lowering terms $\hat\sigma_\pm$ of the drive error that are unaccounted for in the correction protocol. 
We explained in the main text that these are the terms due to the unaccounted axis that persist as a dominant source of infidelity. As such, our case is unlike that considered in Refs.~\cite{Ashhab2007, Lu2012} in which such terms did not exist and the simple gate time correction could be successfully applied.


\section{Infidelity approximations from the Dyson series expansion of time-evolution operators}\label{supp:perturbative expansions}

The fidelity in Eq.~\eqref{eq:fidelity} of the main text is a function in the two-dimensional rescaled parameter space $\tilde\Omega, \tilde\omega_r$ which we express explicitly stage-wise as
\begin{align}
	\mathcal{F}_{\text{c/s}} = \frac{1}{4} \Bigg|\Tr \prescript{}{r}{\bra{0}} \SE(5) \SE(4) \SE(3) \SE(2) \SE(1) \SR(1)^\dagger \SR(2)^\dagger \SR(3)^\dagger \SR(4)^\dagger \SR(5)^\dagger \ket{0}_r \Bigg|. \label{eq:infidelity}
\end{align}
The Dyson series expansions of the time-evolution operators of each stage,
\begin{subequations}
	\begin{align}
		\SR(\stage)^\dagger &= \id + i \underbrace{\int_{\tiz{\stage}}^{\tfz{\stage}} d\tau \HR(\stage)}_{=\opIr{1} (\stage)} - \underbrace{\frac{1}{2} \left(\int_{\tiz{\stage}}^{\tfz{\stage}} d\tau \HR(\stage)\right)^2}_{=\opIr{2} (\stage)} + \dotsi \nonumber \\
		&= \id + i \opIr{1} (\stage) - \opIr{2} (\stage) + \dotsi , \\
		\SE(\stage) &= \id - i \underbrace{\int_{\tiz{\stage}}^{\tfz{\stage}} d\tau \HEsc(\stage; \tau)}_{=\opIe{1} (\stage)} - \underbrace{\int_{\tiz{\stage}}^{\tfz{\stage}} d\tau_1 \int_{\tiz{\stage}}^{\tau_1} d\tau_2 \, \HEsc(\stage; \tau_1) \HEsc(\stage; \tau_2)}_{=\opIe{2} (\stage) }  + \dotsi  \nonumber \\
		&= \id - i \opIe{1} (\stage) - \opIe{2} (\stage)  + \dotsi , \label{eq:exact_evolution_Dyson}
	\end{align}
\end{subequations}
can be decomposed into the operator basis $\{\opvec_\lambda\}$ and its corresponding functions $\Hcoee^{\lambda}$ using the superoperator formalism outlined in the main text. This yields, for the exact Hamiltonian,
\begin{align}
	\HEsc(b;\tau) &= \sum_\lambda\left(\Hcoee^\lambda(b;0) + \sum_{\tilde\omega} \Hcoee^\lambda(b;\tilde\omega) e^{i\tilde\omega \tau} \right)\, \opvec_\lambda \nonumber \\
	&= \sum_\lambda\left( \Hcoer^\lambda(b;0) + \sum_{\tilde\omega} \Hcoee^\lambda(b;\tilde\omega) e^{i\tilde\omega \tau} \right)\, \opvec_\lambda \label{eq:hexact(0)}
\end{align}
where, in the second line, $\Hcoee(\stage; 0) = \Hcoer(\stage; 0)$ is a consequence of the resonance condition and the second sum is over the oscillatory modes in the error terms. We do the same for the red-sideband Hamiltonian, which does not contain time-dependent terms, yielding a much simpler expression,
\begin{align}
	\HR(\stage) &= \sum_\lambda  \Hcoer^\lambda(b;0) \opvec_\lambda. \label{eq:hred(0)}
\end{align}
Therefore, the integrals in the series expansion of the exact evolution can be expressed as
\begin{subequations}
	\begin{align}
		\opIe{1}(\stage) &= \sum_\lambda \opvec_\lambda \int_{\tiz{\stage}}^{\tfz{\stage}} d\tau \, \Hcoee^{\lambda}(\stage;\tau) = \sum_\lambda \opvec_\lambda \Ie{1}{}^\lambda(\stage), \\
		\opIe{2}(\stage) &= \sum_{\lambda \lambda'} \opvec_\lambda \opvec_{\lambda'} \int_{\tiz{\stage}}^{\tfz{\stage}} d\tau_1 \int_{\tiz{\stage}}^{\tau_1} d\tau_2 \, \Hcoee^{\lambda}(\stage;\tau_1) \Hcoee^{\lambda'}(\stage;\tau_2) = \sum_{\lambda \lambda'} \opvec_\lambda \opvec_{\lambda'} \Ie{2}{}^{\lambda \lambda'}(\stage).
	\end{align}
\end{subequations}
The integrals of $\SR(\stage)^\dag$ are similarly decomposed. 
The general form of the one- and two-dimensional integrals are
\begin{subequations}
	\begin{align}
		\Ir{1}{}^{\lambda}(b) = &\Hcoer^\lambda(\stage; 0) \Delta\tau_\stage \\
		\Ir{2}{}^{\lambda \lambda'}(\stage) = &\frac{\Delta\tau_\stage^2}{2} \Hcoer^\lambda(\stage; 0) \Hcoer^{\lambda'}(\stage; 0) = \frac{1}{2} \Ir{1}{}^{\lambda}(\stage) \Ir{1}{}^{\lambda'}(\stage)\\
		\Ie{1}{}^\lambda(\stage) = &\Hcoee^\lambda(\stage; 0) \Delta\tau_\stage + \sum_{\tilde\omega} \frac{\Hcoee^\lambda(\stage; \tilde\omega) e^{i \tilde\omega \tau}}{i \tilde\omega} \Big|_{\tiz{\stage}}^{\tfz{\stage}} \\
		\Ie{2}{}^{\lambda \lambda'}(\stage) = &\frac{\Delta\tau_\stage^2}{2} \Hcoee^\lambda(\stage; 0) \Hcoee^{\lambda'}(\stage; 0) \nonumber \\
		+ &\Hcoee^\lambda(\stage; 0) \sum_{\tilde\omega'} \left(\frac{\Hcoee^{\lambda'}(\stage; \tilde\omega') e^{i\tilde\omega'\tau}}{(i\tilde\omega')^2}\Big|_{\tiz{\stage}}^{\tfz{\stage}} - \frac{\Hcoee^{\lambda'}(\stage; \tilde\omega') e^{i\tilde\omega'\tiz{\stage}} \Delta\tau_\stage}{i\tilde\omega'}\right) \nonumber \\
		+ \sum_{\tilde\omega} &\left(\frac{\Hcoee^\lambda(\stage; \tilde\omega) e^{i\tilde\omega\tau}}{\tilde\omega^2}\Big|_{\tiz{\stage}}^{\tfz{\stage}} + \frac{\Hcoee^\lambda(\stage; \tilde\omega) e^{i\tilde\omega\tfz{\stage}} \Delta\tau_\stage}{i\tilde\omega}\right) \Hcoee^{\lambda'}(\stage; 0) \nonumber \\
		+ \sum_{\tilde\omega \tilde\omega'} \frac{1}{i\tilde\omega'} &\left(\frac{\Hcoee^{\lambda}(\stage; \tilde\omega) e^{i\tilde\omega\tau} \Hcoee^{\lambda'}(\stage; \tilde\omega') e^{i\tilde\omega'\tau}}{i(\tilde\omega + \tilde\omega')}\Big|_{\tiz{\stage}}^{\tfz{\stage}} - \frac{\Hcoee^{\lambda}(\stage; \tilde\omega) e^{i\tilde\omega\tau} \Hcoee^{\lambda'}(\stage; \tilde\omega') e^{i\tilde\omega'\tiz{\stage}}}{i\tilde\omega}\Big|_{\tiz{\stage}}^{\tfz{\stage}}\right) \label{eq:general integrals 2e}
	\end{align}  \label{eq:general integrals}
\end{subequations}

We make three observations here. Firstly, in the calculation of fidelity, the product of the identity terms of all stages in the series expansion of time-evolution yield a trace of 4, which when multiplied with the prefactor of $\frac{1}{4}$, yields unity and allows us to express the fidelity as $\mathcal{F}_{\text{c/s}} = \abs{1 - \delta_{\text{c/s}}}$, where infidelity $\delta_{\text{c/s}}(\tilde\Omega, \tilde\omega_r) \in \mathbbm{C}$ as defined, is a complex number and consists of all higher dimensional $n \ge 1$ integrals. Thus, if Dyson series converges sufficiently fast, we can approximate the infidelity (hence, the fidelity) to the leading order perturbatively. 

Secondly, the consequence of $\Hcoee(\stage; 0) = \Hcoer(\stage; 0)$ is that in the series expansion of time-evolution, for every stage, the exact Hamiltonian contributes $n$-th order terms $\Pi_\lambda \Hcoer^\lambda(\stage; 0) (\Delta \tau_b)^n$ that are exactly equal to the same terms from the ideal red-sideband, to all orders. Thus, these terms cancel exactly in the calculation of infidelity, yielding leftover terms that reflect the deviation of the exact from the ideal evolution. The infidelity series is thus
\begin{align}
	\delta_{\text{c/s}}(\tilde\Omega, \tilde\omega_r) = \sum_n \delta_{\text{c/s}}^{(n)}(\tilde\Omega, \tilde\omega_r) = \sum_n \frac{1}{4}  \Tr\prescript{}{r}{\bra{0}} \text{$n$-th order integrals} \ket{0}_r \label{eq:infidelity series}
\end{align}
where the order $n$ refers to the total dimension of the integral domain.

Thirdly, when taking the trace over two-qubit states and the resonator vacuum state, first order terms in the series expansion of the exact evolution Eq.~\eqref{eq:exact_evolution_Dyson} vanish. The reason is that these contain operators in the error terms that yield zero trace. Thus, the leading order in the infidelity series is the second order, which is given by 
\begin{align}
	&\delta_\text{c/s}^{(2)} \nonumber \\
	&= \frac{1}{4}\Tr\prescript{}{r}{\bra{0}}  \left[ \sum_b \opIe{2}(b) + \sum_b \opIr{2}(b)+ \sum_{b < b'} \opIr{1}(b) \opIr{1}(b') + \sum_{b > b'} \opIe{1}(b) \opIe{1}(b') - \sum_{b, b'} \opIe{1}(b) \opIr{1}(b') \right ]\ket{0}_r. \label{eq:infidelity series2}
\end{align} 
To make our calculations concrete, we illustrate the specific form of $\delta_\text{c/s}^{(2)}$, by employing the following abbreviated notations with the understanding that (i) the first integral on the left is with respect to $d\tau_1$ and the second $d\tau_2$, (ii) where two operators are multiplied, the one on the left has a $\tau_1$ dependence while the one on the right has a $\tau_2$ dependence, unless the operator is the red-sideband Hamiltonian, in which case it is time-independent, and (iii) $\hat V_{\text{c/s}}$ is the sum of all errors, i.e.  $\hat V_{\text{c/s}} \equiv \sum_k \hat V_{\text{c/s},k}$, where $k \in \{\text{red, blue, long, drive} \}$.
\begin{subequations}
	\begin{align}
		\opIr{1}(b)\opIr{1}(b') &\rightarrow \iint \HR(b)\HR(b') \\
		\opIe{1}(b)\opIe{1}(b') &\rightarrow \iint \left(\HR(b)\HR(b') + \HR(b) \hat V_{\text{c/s}}(b') + \hat V_{\text{c/s}}(b) \HR(b') + \hat V_{\text{c/s}}(b) \hat V_{\text{c/s}}(b'))\right) \\
		\opIe{1}(b)\opIr{1}(b') &\rightarrow \iint \left(\HR(b) \HR(b') + \hat V_{\text{c/s}}(b) \HR(b') \right)\\
		\opIr{2}(b) &\rightarrow \iint \frac{1}{2}\HR(b)\HR(b) \\
		\opIe{2}(b) &\rightarrow \iint \left(\frac{1}{2}\HR(b)\HR(b) + \mathcal{T}(\HR(b) \hat V_{\text{c/s}}(b) + \hat V_{\text{c/s}}(b) \HR(b) + \hat V_{\text{c/s}}(b) \hat V_{\text{c/s}}(b))\right) \; .
	\end{align}
\end{subequations}
Decomposing the summation $\sum_{b, b'} = \sum_{b > b'} + \sum_{b < b'} + \sum_{b=b'}$, we obtain the form of the second-order infidelity given in the main text,
\begin{align}
	\delta_\text{c/s}^{(2)} &= \frac{1}{4} \iint \Tr \prescript{}{r}{\bra{0}} \Big(\sum_{b<b'} \hat V_{\text{c/s}}(b) \HR(b') + \sum_{b>b'} (\HR(b)\hat V_{\text{c/s}}(b') + \hat V_{\text{c/s}}(b) \hat V_{\text{c/s}}(b')) \nonumber \\
	&+ \sum_{b} (\hat V_{\text{c/s}}(b) \HR(b) - \mathcal{T}(\{\HR(b), \hat V_{\text{c/s}}(b)\} + \hat V_{\text{c/s}}(b) \hat V_{\text{c/s}}(b)))\Big) \ket{0}_r. \label{eq:infidelity series3}
\end{align}
We observe that all terms that are quadratic in $\HR$ cancel, consistent with the discussion earlier. Inspecting the surviving terms, we see that infidelity arises from the cross terms between the red sideband and deviations, terms that are quadratic in the deviations, and the effect of time ordering of non-commutative error terms.


\subsection{Infidelity convergence} \label{sec:Infidelity convergence}

In this section, we justify our claim made in the main text that, if the second-order infidelity function satisfies $\abs{\delta_\text{c/s}^{(2)}(\tilde\Omega, \tilde\omega_r)} < 1/4$, the infidelity forms a convergent and perturbative series. The preimage of the leading order $\abs{\delta_\text{c/s}^{(2)}}$ therefore defines the high-fidelity regimes in the parameter space.

We check the convergence of the series Eq.~\eqref{eq:infidelity series}. Consider first, the real function
\begin{align}
	Q_1(\tilde\omega) := \abs{\int_{\tiz{\stage}}^{\tfz{\stage}} d\tau \exp(i \tilde\omega \tau)} = 
	\begin{cases}
		\frac{1}{\abs{\tilde\omega}}\abs{e^{i \tilde\omega \tfz{\stage}} - e^{i \tilde\omega \tiz{\stage}}} \;\;\;\; \tilde\omega \in \mathbbm{R}\setminus\{0\} \\
		\Delta \tau_\stage \;\;\;\; \omega \in \{0\} 
	\end{cases}
\end{align}
wherein the singularity at $\omega=0$ is a removable singularity where the function takes its maximum and approaches $0$ asymptotically. This is the integral we encountered in the first order (1d) integral of Dyson series. These properties can be generalized to higher dimensional integrals, and are responsible for the spurious resonant frequencies responsible for fidelity losses identified in the main text.

From the above, we see that the $n$-th order integral of the infidelity function would be maximum if the Hamiltonian in the integrand were a constant, which is not possible for our case since the exact Hamiltonians contain time-dependent error terms. What this observation allows us to do is to identify an upper bound arising from for a constant Hamiltonian, where if the summation of all terms with non-zero trace yields a convergent infidelity series, the series expansion is legitimate. We can then satisfy the inequality
\begin{align}
	\abs{4\delta_\text{c/s}^{(n)}} \leq \sum_{\{\lambda\}} \Tr \prescript{}{r}{\bra{0}} \opvec^{\lambda_1} \cdots \opvec^{\lambda_n}\ket{0}_r \frac{r^n}{n!} < \sum_{\{\lambda\}} \Tr \prescript{}{r}{\bra{0}} \opvec^{\lambda_1} \cdots \opvec^{\lambda_n}\ket{0}_r \frac{\Delta\tau_{\text{total}}^n}{n!}
\end{align}
where $\sum_{\{\lambda\}}$ refers to the sum of terms with non-zero trace, and $r^n/n!$ is some intermediate value between the maximum $\Delta\tau_{\text{total}}^n / n!$ and the actual $n$-th order modulus. It is this intermediate value $r$ which gives a tighter bound, that we aim to deduce.

Each operator basis element $\opvec^{\lambda}$ lives in the Hilbert space of the two-qubit and photon states. For the qubit subspace, Hermitian and unitary generators (Pauli matrices) preserve the trace. However, products of creation and annihilation operators in the photon subspace result in terms which, after tracing over the photon vacuum state, are 
\begin{align}
	\prescript{}{r}{\bra{0}}  \a^m \c{}^n \ket{0}_r = 
	\begin{cases}
		0 \;\;\; m \neq n \\
		n! \;\;\; m = n
	\end{cases} .
\end{align}
Now, we perform a combinatoric analysis of the terms. If the Hamiltonian contains free qubit terms $\Xj + \Yj + \Zj$, the operator polynomial at $n$-th order is $(\Xj + \Yj + \Zj)^n$ contains $3^n$ terms, some of which are zero after taking the trace. For qubit-photon interaction with the form $(\Xj + \Yj + \Zj)(\a + \c)$, only even orders where $n=2m$ contribute terms with non-zero trace. For example, $\prescript{}{r}{\bra{0}}(\a + \c)^{2m}\ket{0}_r$ contains $C^{2m-2}_{m-1}$ possible non-zero combination of $\a$'s and $\c$'s, wherein: \\ \\
$\prescript{}{r}{\bra{0}}\a^m \c{}^m\ket{0}_r$ = $m!$ \\
$\prescript{}{r}{\bra{0}}\a^{m-1} \c \a \c{}^{m-1}\ket{0}_r$ = $m! - (m-1)!$ \\
$\prescript{}{r}{\bra{0}}\a^{m-2} \c \a \c \a \c{}^{m-2}\ket{0}_r = m! - 3(m-1)! + (m-2)!$ \\
$\prescript{}{r}{\bra{0}}\a^{m-2} \c \c \a \a \c{}^{m-2}\ket{0}_r = m! - 4(m-1)! + 2(m-2)!$ \\
$\cdots$ \\

We see that every combination contributes a value of $m!$, every commutation between $\a$ and $\c$ contributes a value of $(m-1)!$, and the position of such a commutation contributes an additional value of $(m-p)!$ where $p \in \{p \in \mathbbm{N}; p < m-1\}$, and so on.  By counting the terms and collecting the terms with the largest possible magnitude, we have
\begin{subequations}
	\begin{align}
		\abs{4\delta_\text{c/s}^{(2m)}} < N(r,2m) &:= \frac{4 (3r)^{2m}}{(2m)!} \left(\frac{m!(2m-2)!}{(m-1)!^2} - \mathcal{O}((m-1)!)\right) \label{eq:2m estimate} \\
		\abs{4\delta_\text{c/s}^{(2m+1)}} < N(r,2m+1) &:= \frac{ 4 (3r)^{2m+1}}{(2m+1)!} \label{eq:2m+1 estimate}
	\end{align}
\end{subequations}
where the prefactor $4$ on the right-hand side comes from the trace of the two-qubit identity operator, i.e., we assume that among $3^n$ terms of the two-qubit trace, every one of them contributes a trace $4$, which is strictly larger than the actual integrals (the series converges regardless).

This assures the absolute convergence of the infidelity series by checking Eq.~\eqref{eq:2m estimate} and~\eqref{eq:2m+1 estimate} at $m \rightarrow \infty$. Since $C^{2m-2}_{m-1} m!$ grows faster than $(m-1)!, (m-2)!, \cdots$ with their combinatorics for all $m$, we can safely ignore them when analyzing the general behavior of the order of magnitude of the $2m$-th order integral. For $(2m+1)$-th order integral, the free qubit terms make a smaller contribution after the trace operation than that of the $2m$-th order terms because of the properties of $SU(2)$ generators. There are no Fock space contributions of the trace in the $(2m+1)$-th order integral, so it grows significantly slower than that of the $2m$-th order. Additionally, all first order integrals contain operators that vanish after tracing over the two-qubit and resonator vacuum states, so they do not contribute. Therefore, we focus on the magnitude of $2m$-th order integral. 
By an analytic continuation from $\mathbbm{N}$ to $\mathbbm{R}_{+}$, we first plot Eq.~\eqref{eq:2m estimate} with $r \in [1/12, 7/12]$ in Supp. Fig.~\ref{fig: modules}.

\begin{figure}
	\centering
	\includegraphics[width=0.5\linewidth]{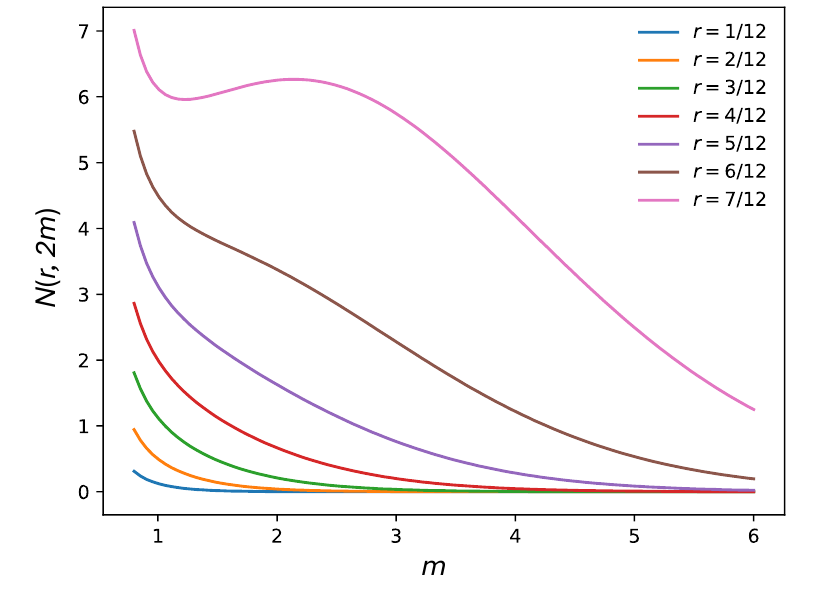}
	\caption{Upper bound $N(r, 2m)$ of the $2m$-th term in the infidelity expansion against the order $m$ of the terms, for various values of $r$. Convergence requires monotonically decreasing values as the order $m$ increases. Here, with the exception of $r=7/12$ (pink line), all other values ensure convergence, enabling the set of $r$ values in Eq.~\eqref{eq:r_values} to be obtained.}
	\label{fig: modules}
\end{figure}

Crucially, our goal is to find a suitable value of $r$ so that the infidelity function can be perturbatively expanded, which is equivalent to find a suitable value of $r$ that renders the modulus monotonically decreasing. In the domain of interest $m \in [1, \infty)$, $1/(2m)! \rightarrow 1/\Gamma(2m+1)$ oppresses the overall magnitude and ensure convergence. However, $(m-1)! \rightarrow \Gamma(m)$ is not a monotonic function in this domain, specifically in the subset $[1,2]$. 

We determine $r$ by requiring that the derivative $N'(r,2m) := \partial_m N(r,2m)$ for $m \in [1,2]$ has to be negative for that value of $r$. This function can be expressed in terms of zeroth order polygamma functions $\psi^{(0)}(z) := \partial_z \text{Log} \, \Gamma(z), z \in \mathbbm{C}$. A set of $r$ values can be found by requiring that the derivative of $N(r,2m)$ at $m=3/2$ is $-1$, i.e. 
\begin{align}
	D(r) := \{r \in \mathbbm{R}_{+} \; ; \; N'(r,3) = -1\}, \text{ for which $m=3/2 \in [1,2]$}. \label{eq:r_values}
\end{align}
Thus, $\min D(r)$ gives the suitable $r$ for the infidelity series with monotonically decreasing terms, i.e. a convergent series. We find $\min D(r) := r_{\text{mono}} \approx 0.295$, which gives $N(r_{\text{mono}}, 2) \approx 1.563$ and $N(r_{\text{mono}}, 4) \approx 0.407$. Thus, we safely claim that in the high-fidelity regimes $F_{\text{high}}$ defined by the preimage in the domain of the two-dimensional rescaled parameter space $(\tilde\Omega, \tilde\omega_r)$, 
\begin{align}
	F_{\text{high}} := \Big\{(\tilde\Omega, \tilde\omega_r) \in \mathbbm{R}^2; \; \abs{4\delta_\text{c/s}^{(2)}} < 1 < N(r_{\text{mono}}, 2) \Big\},
\end{align}
i.e. the infidelity series is well approximated by $\abs{\delta_\text{c/s}^{(2)}} < 1/4$, as mentioned in the main text.

From the discussion above, we understand the order of magnitude of the $n$-th order series expansion of the infidelity: The high-fidelity regions in the parameter space are located at the space where the series converges fast enough, i.e., the second order provides a perturbative approximation of the infidelity; the low-fidelity regions (the complement of the high-fidelity regions) correspond to a large modulus (related to $r$) of each term in the series. There are two types of low-fidelity regions, which can be explained from a physical point of view:
\begin{enumerate}
	\item Around the neighborhood of the removable singularities $\Omega_s$ with radius $\eta(\Omega_s)$ in the parameter space, denoted by $B(\eta(\Omega_s), \Omega_s)$, where some time-dependent terms except the red sideband transition in the exact Hamiltonian become constant, the time evolution consists of both red sideband transitions and other evolution generated by unwanted constant Hamiltonians (generally non-commutative with the desired evolution) with comparable magnitude. Their contribution is proportional to the total gate time $\Delta\tau_{\text{total}}$, which is larger than $1$. 
	\item  In the region where the parameter dependent magnitude, e.g., the magnitude of the free qubit Hamiltonian of the spin qubit is proportional to $\Omega$, is large enough, the terms other than the desired red sideband transition dominant, contribute to a large infidelity and the series does not converge.
\end{enumerate}

We are now ready to calculate the general integrals Eq.~\eqref{eq:general integrals} and substitute them into Eq.~\eqref{eq:infidelity series3}, under the assumption that $(\tilde\Omega, \tilde\omega_r) \in F_{\text{high}}$. In the following two subsections, we calculate explicitly the map of high-fidelity regions.

\subsection{Around the removable singularities}

The frequencies at the removable singularities are responsible for the spurious resonances as described in the main text. If we fix $\tilde\omega_r$, both spin and charge qubit system have a removable singularity at $\tilde\Omega = 0$. For the spin qubit, we evaluate the second order integrals in Eq.~\eqref{eq:general integrals} involving $\{\tilde\omega\} = \{\pm 2\tilde\Omega, \pm 4\tilde\Omega\}$, to obtain
\begin{align}
	4\delta_\text{s}^{(2)} \approx \sum_{b=1}^{5} \sum_{\text{non-zero trace} \{\lambda\}} \Ies{2}{}^{\lambda_1 \lambda_2}(b; \pm 2\tilde\Omega, \pm 4\tilde\Omega) =& \frac{1}{4} \left(\frac{4}{i 2\tilde\Omega} + \frac{2}{i 4\tilde\Omega}\right)\Delta\tau_{\text{total}} + \mathcal{O}(\tilde\Omega^{-2}, \tilde\omega_d^{-1}) \nonumber \\
	=& \frac{1}{4} \left(\frac{5}{i 2\tilde\Omega}\right)\Delta\tau_{\text{total}},
\end{align}
where $\Delta\tau_{\text{total}} = \sum_{b=1}^5 \Delta\tau_b = (3 + \sqrt{2})\pi$ is the total dimensionless gate time. 
The notation $\mathcal{O}(\tilde\omega^{-n})$ denotes terms that are asymptotically similar to the Fourier mode-expansion of the form $\tilde\omega^{-n} \exp(i\tilde\omega' \tau)$ for some $\tilde\omega'$ that are allowed to be zero. The amplitude is singled out since the oscillatory function $\exp(i\tilde\omega' \tau)$ is bounded when $\tilde\omega', \tau \in \mathbbm{R}$, while $\tilde\omega^{-n}, \tilde\omega \in \mathbbm{R}$ is not bounded.
The prefactor $1/4$ comes from the Fourier complex coefficient $1/2i$ of the sine function. 
The multiplication of two error terms $\Tr \prescript{}{r}{\bra{0}}\iint \hat V(b) \hat V(b') \ket{0}_r$ , on the other hand, contributes $\mathcal{O}(\tilde\Omega^{-2})$, and is therefore negligible in high-fidelity regimes.
Additionally, the cross term between red sideband and errors $\Tr \prescript{}{r}{\bra{0}}\iint \mathcal{T}(\{\HR(b), \hat V(b)\} \ket{0}_r$ and $\Tr \prescript{}{r}{\bra{0}}\iint \HR(b)\hat V(b') \ket{0}_r$ contribute $\mathcal{O}(\tilde\omega_d^{-1}) = \mathcal{O}((\tilde\omega_r - 2\tilde\Omega)^{-1})$, which is also negligible around $\tilde\Omega = 0$ given that $\tilde\omega_r \gg 1$.

The terms that contribute are the oscillating functions with frequencies $\pm 4\tilde\Omega$ coming from $\bra{0}\a \mj \c \pj\ket{0}$ (the blue sideband terms) with trace $2$, and those with frequencies $\pm \tilde2\Omega$ coming from $\bra{0}\a \Zj \c \Zj\ket{0}$ with trace $4$ (longitudinal interaction terms). Since we have established that in the high-fidelity regimes, $\abs{4\delta_\text{c/s}^{(2)}} < 1$, we thus obtain the condition
\begin{align}
	\abs{2\tilde\Omega} > \frac{5}{4} \Delta\tau_{\text{total}},
\end{align}
neglecting corrections of order $\mathcal{O}(\tilde\Omega^{-2}, \tilde\omega_d^{-1})$.

For charge qubit, the second order integrals with frequencies $\pm 4\tilde\Omega$ to the leading order gives the same results as that of the spin qubit: $\frac{2}{4} \frac{\Delta\tau_\text{total}}{i4\Omega}$. 
The frequencies $\pm 2\tilde\Omega$ contributed by the longitudinal error term have a more complicated form because of the detuning drive, which contains a square root of an oscillating function. At $\tilde\Omega = 0$, the integral over 
\begin{align}
	2\xi_1 \cos(\tilde\omega_d \tau + \phi) \Zj \left(\a e^{-i(2\tilde\Omega +\tilde\omega_d) \tau} + \c e^{i (2\tilde\Omega +\tilde\omega_d) \tau}\right)
\end{align}
contributes to a removable singularity. Take  the trace $\Tr \prescript{}{r}{\bra{0}}\a \Zj \c \Zj\ket{0}_r = 4$ into account, the second order integrals with $\omega = \pm 4\tilde\Omega, \pm 2\tilde\Omega$ yields
\begin{align}
	\begin{split}
		4\delta_\text{c}^{(2)} \approx \sum_{b=1}^{5} \sum_{\text{non-zero trace} \{\lambda\}} \Iec{2}{}^{\lambda_1 \lambda_2}(b; \pm 2\tilde\Omega, \pm 4\tilde\Omega) &= \left(\frac{1}{4} \frac{2}{i4\tilde\Omega} + 4 (2\xi_1)^2 \frac{1}{i2\tilde\Omega}\right) \Delta\tau_\text{total} + \mathcal{O}(\tilde\Omega^{-2}, \tilde\omega_d^{-1})  \\
		&\approx \frac{\Delta\tau_\text{total}}{i2\tilde\Omega}\left(\frac{1}{4} + 4 (2\xi_1)^2\right)
	\end{split}
\end{align}
The high-fidelity region of the charge qubit  thus satisfies
\begin{align}\label{eq:charge qubit widths}
	\abs{2\tilde\Omega} > \left(\frac{1}{4} + 4 (2\xi_1)^2\right) \Delta\tau_\text{total} \approx 0.865 \Delta\tau_\text{total} 
\end{align}

For charge qubit, there are many removable singularities correspond to each $n \geq 1$ in Eq.~\eqref{eq:epsilon fourier cosine} located at $2\tilde\Omega = \tilde\omega_r\left(1 - \frac{1}{n}\right)$. Analogous to the previous analysis at $\Omega = 0$ ($n = 1$), we argue that for all $1 < n < \infty$, the high-fidelity regimes additionally satisfies
\begin{align}
	\abs{2\tilde\Omega - \tilde\omega_r\left(1 - \frac{1}{n}\right)} > 4(2\xi_n)^2 \Delta\tau_\text{total}.
\end{align}
According to the calculation in Eq.~\eqref{eq: sqrt wave coefficient}, the width of low-fidelity regimes is decreasing with an increasing $n$.

When $2\tilde\Omega \rightarrow \tilde\omega_r$ ($n=\infty$), we have $\tilde\omega_d \rightarrow 0$. 
Compared to other removable singularities that leads to an constant term in the longitudinal interaction with small amplitude, the frequency $2\tilde\Omega = \tilde\omega_r$ renders the entire red-sideband error term $\hat V_{\text{red}, t_c}^{(j)}(t)$ to become a constant. In addition, when $\phi = 0$, $\hat V_{\text{red}, t_c}^{(j)}(t)$ cancels exactly with $\HR^{(j)}$. 
The non-zero trace terms includes $\Tr \bra{0}\a \pj \c \mj\ket{0} = 2$ such that, to the leading order,
\begin{align}
	4\delta_\text{c}^{(2)} \approx \sum_{b=1}^{5} \sum_{\text{non-zero trace} \{\lambda\}} \Iec{2}{}^{\lambda_1 \lambda_2}(b; \pm 2\tilde\Omega, \pm 4\tilde\Omega) &= \frac{1}{4} \left(\frac{2}{i2\tilde\omega_d} + \frac{8}{i\tilde\omega_d} \right)\Delta\tau_\text{total} + \mathcal{O}(\tilde\omega_d^{-2}, \tilde\Omega^{-1}, \Delta\tau_b \tilde\omega_d^{-1}) \nonumber \\
	&  \approx \frac{\Delta\tau_\text{total}}{i\tilde\omega_d} \frac{9}{4}.
\end{align}
The cross terms $\iint \Tr\prescript{}{r}{\bra{0}} \hat V(b) \HR(b') \ket{0}_r$, $\iint \Tr\prescript{}{r}{\bra{0}} \HR(b
) \hat V(b')  \ket{0}_r$, and the term $\iint \Tr\prescript{}{r}{\bra{0}}  \hat V(b) \hat V(b')\ket{0}_r$ contribute terms of order $\mathcal{O}(\Delta\tau_b \tilde\omega_d^{-1})$ and $\mathcal{O}(\tilde\omega_d^{-2}, \tilde\Omega^{-1})$ respectively. 
We solve the high-fidelity condition around $2\tilde\Omega \rightarrow \tilde\omega_r$ to the leading order, yielding
\begin{align}
	\abs{2\tilde\Omega - \tilde\omega_r} >  \frac{9}{4} \Delta\tau_\text{total}.
\end{align}

The result above yields corrections from the next leading order $\mathcal{O}(\Delta\tau_b \tilde\omega_d^{-1}))$ which is small after summing all contributions. 
This comes from the following observation: the magnitude of each these terms in $\mathcal{O}(\Delta\tau_b \tilde\omega_d^{-1}))$ is smaller or equal to $\frac{4\sqrt{2}}{9(3 + \sqrt{2})} \approx 0.14$. 
The cross terms $\sum_{b} \iint \hat V(b) \HR(b)$ contains $5$ terms which are equal to $\frac{4\sqrt{2}}{9(3 + \sqrt{2})}$, while the cross terms between different stages yields $8$ terms with non-zero trace that are equal to $\frac{4\sqrt{2}}{9(3 + \sqrt{2})}$. The summation of all of these terms yields an estimated amplitude $13\frac{4\sqrt{2}}{9(3 + \sqrt{2})} \approx 1.9$ which is small compared to the result above, $\frac{9}{4} \Delta\tau_\text{total} = \frac{9}{4}(3+\sqrt{2}\pi) \approx 16.75$.

\subsection{Spin qubit upper bounds}

Now we explicitly detail the steps leading to the upper bound on drive amplitude for the spin qubit, shown in Eq.~\eqref{eq:bounds spin qubit} in the main text. 

For the spin qubit, the upper bound of the drive amplitude arises from the drive error term $\hat V_\text{s,drive}(t)$, whose amplitude is proportional to parameter $\tilde\Omega$ (corresponds to the second type \textit{(2)} of low-fidelity region described at the end of Sec.~\ref{sec:Infidelity convergence}). Within this error term, we pay attention to the operators that yield non-zero trace and thus contribute to infidelity. These are the terms $\pj \mj$, $\mj \pj$, and $\Zj \Zj$. The contribution to infidelity from these terms are then
\begin{subequations}
	\begin{align}
		&\sum_{b=1}^{5} \sum_{\text{non-zero trace} \{\lambda\}} \Ies{2}{}^{\lambda_1 \lambda_2}(b) = 4 \Re \Ies{2}{}^{12}(b) + 4 \Ies{2}{}^{33}(b) \, \\
		&\sum_{b > b'=1}^{5} \sum_{\text{non-zero trace} \{\lambda\}} \Ies{1}{}^{\lambda_1}(b) \Ies{1}{}^{\lambda_2}(b') \nonumber \\
		&= \sum_{b > b'=1}^{5}\left(2\Ies{1}{}^{1}(b) \Ies{1}{}^{2}(b') + 2\Ies{1}{}^{2}(b) \Ies{1}{}^{1}(b') + 4\Ies{1}{}^{3}(b) \Ies{1}{}^{3}(b')\right) 
	\end{align}
\end{subequations}

The constituent integrals are explicitly,
\begin{subequations}
	\begin{align}
		\begin{split}
			\Ies{2}{}^{12}(b) =  &\frac{i\tilde\Omega^3}{2(\tilde\omega_d^2 - \tilde\Omega^2)}\left(-\frac{\sin(4\tilde\omega \tau + 4\phi)}{8\tilde\omega_d} + \frac{\tau}{2}\right)\eval^{\tfz{\stage}}_{\tiz{\stage}} + \frac{\tilde\Omega^2 \tilde\omega_d}{16\tilde\omega_d(\tilde\omega_d^2 - \tilde\Omega_d^2)}\cos(4\tilde\omega_d \tau + 4\phi)\eval^{\tfz{\stage}}_{\tiz{\stage}} \\
			&+ \frac{\tilde\Omega^2}{4(\tilde\omega_d^2 - \tilde\Omega^2)^2} \left(e^{2i\tilde\Omega \tau}\left(i\tilde\Omega \sin(2\tilde\omega_d\tau + 2\phi) - \tilde\omega_d \cos(2\tilde\omega_d\tau + 2\phi)\right)\right)\eval^{\tfz{\stage}}_{\tiz{\stage}} \\
			&\times e^{-2i\tilde\Omega \tiz{\stage}}\left(i\tilde\Omega \sin(2\tilde\omega_d \tiz{\stage} + 2\phi) + \tilde\omega_d \cos(2\tilde\omega_d \tiz{\stage} + 2\phi)\right)
		\end{split} \\
		\Ies{2}{}^{33}(b) &= -\frac{\tilde\Omega^2}{16\tilde\omega_d^2}\cos(4\tilde\omega_d \tau + 4\phi)\eval^{\tfz{\stage}}_{\tiz{\stage}} - \frac{\tilde\Omega^2}{4\tilde\omega_d^2} \sin(2\tilde\omega_d \tiz{\stage} + 2\phi) \sin(2\tilde\omega_d \tau + 2\phi)\eval^{\tfz{\stage}}_{\tiz{\stage}} \\
		\Ies{1}{}^{1}(b) &= \frac{e^{2i\tilde\Omega\tau} \left(i\tilde\Omega^2 \sin\left(2\tilde\omega_d \tau + 2{\phi}\right) - \tilde\Omega \tilde\omega_d \cos\left(2\tilde\omega_d \tau + 2{\phi}\right)\right)}{2 \left(\tilde\omega_d^2 - \tilde\Omega^2\right)}\eval^{\tfz{\stage}}_{\tiz{\stage}} \\
		\Ies{1}{}^{2}(b) &= \frac{e^{-2i\tilde\Omega\tau} \left(-i\tilde\Omega^2 \sin\left(2\tilde\omega_d \tau + 2{\phi}\right) - \tilde\Omega \tilde\omega_d \cos\left(2\tilde\omega_d \tau + 2{\phi}\right)\right)}{2 \left(\tilde\omega_d^2 - \tilde\Omega^2\right)}\eval^{\tfz{\stage}}_{\tiz{\stage}} \\
		\Ies{1}{}^{3}(b) &= \tilde\Omega\frac{\sin\left(2\tilde\omega_d \tau + 2{\phi}\right)}{2\tilde\omega_d}\eval^{\tfz{\stage}}_{\tiz{\stage}}
	\end{align}
\end{subequations}
Here, the superscripts $\lambda_1, \lambda_2$ serve as book-keeping labels for our notation. To maintain the flow of the argument, we omit the detailed explanation of these indices. Continuing, we can write
\begin{subequations}
	\begin{align}
		\sum_{b=1}^{5} \sum_{\text{non-zero trace} \{\lambda\}} \Ies{2}{}^{\lambda_1 \lambda_2}(b) &=  \sum_{i} A_i \, f_i(b, \tau)\eval^{\tfz{\stage}}_{\tiz{\stage}} \\
		\sum_{b > b'=1}^{5} \sum_{\text{non-zero trace} \{\lambda\}} \Ies{1}{}^{\lambda_1}(b) \Ies{1}{}^{\lambda_2}(b') &= \sum_{ij} B_{ij} \, f_i(b, \tau)\eval^{\tfz{\stage}}_{\tiz{\stage}} f_j(b', \tau)\eval^{\tfz{\stage'}}_{\tiz{\stage'}}
	\end{align}
\end{subequations}

where
\begin{subequations}
	\begin{align}
		(A_i) &= \left(-\frac{\tilde\Omega^2}{4\tilde\omega_d^2}, -\frac{\tilde\Omega^2}{\tilde\omega_d^2}, -\frac{\tilde\Omega^2}{4(\tilde\omega_d^2 - \tilde\Omega^2)}, -\frac{\tilde\Omega^4}{(\tilde\omega_d^2 - \tilde\Omega^2)^2}, -\frac{\tilde\Omega^3 \tilde\omega_d}{(\tilde\omega_d^2 - \tilde\Omega^2)^2}, + \frac{\tilde\Omega^3 \tilde\omega_d}{(\tilde\omega_d^2 - \tilde\Omega^2)^2}, -\frac{\tilde\Omega^2 \tilde\omega_d^2}{(\tilde\omega_d^2 - \tilde\Omega^2)^2}\right)  \label{eq:Ai} \\
		(B_{ij}) &= \Big(\frac{M}{2} \otimes \begin{pmatrix}
			\frac{\tilde\Omega^4}{\left(\tilde\omega_d^2 - \tilde\Omega^2\right)^2} & -i\frac{\tilde\Omega^3 \tilde\omega_d}{\left(\tilde\omega_d^2 - \tilde\Omega^2\right)^2} \\
			+i \frac{\tilde\Omega^3 \tilde\omega_d}{\left(\tilde\omega_d^2 - \tilde\Omega^2\right)^2} & \frac{\tilde\Omega^2 \tilde\omega_d^2}{\left(\tilde\omega_d^2 - \tilde\Omega^2\right)^2}
		\end{pmatrix} \Big)_{\substack{(\lambda_1, \lambda_2) \\ = (1,2)}} \oplus \Big(\frac{M}{2} \otimes
		\begin{pmatrix}
			\frac{\tilde\Omega^4}{\left(\tilde\omega_d^2 - \tilde\Omega^2\right)^2} & -i\frac{\tilde\Omega^3 \tilde\omega_d}{\left(\tilde\omega_d^2 - \tilde\Omega^2\right)^2}  \\
			+i \frac{\tilde\Omega^3 \tilde\omega_d}{\left(\tilde\omega_d^2 - \tilde\Omega^2\right)^2} & \frac{\tilde\Omega^2 \tilde\omega_d^2}{\left(\tilde\omega_d^2 - \tilde\Omega^2\right)^2}
		\end{pmatrix} \Big)_{\substack{(\lambda_1, \lambda_2) \\ = (2,1)}} \oplus 
		\begin{pmatrix}
			\frac{\tilde\Omega^2}{\tilde\omega_d^2}
		\end{pmatrix}_{\substack{(\lambda_1, \lambda_2) \\ = (3,3)}} \label{eq:Bij}
	\end{align}
\end{subequations}
and $f_i(b, \tau)$ that satisfy $f_i(b, \mathbbm{R}) \in [-1, +1]$ and are some periodic functions composed by sine and cosine at stage $b$. For example, $+ \frac{\tilde\Omega^3 \tilde\omega_d}{(\tilde\omega_d^2 - \tilde\Omega^2)^2}$ in $A$ corresponds to $\sin(2\tilde\Omega \tau - 2\tilde\Omega \tiz{\stage}) \cos(2\tilde\omega_d \tau + 2\phi) \sin(2\tilde\omega_d \tiz{\stage} + 2\phi)\eval^{\tfz{\stage}}_{\tiz{\stage}}$. 
To count the amplitudes of each terms of periodic functions $f_i(\stage, \tau)$ in $(B_{ij})$, we expand $\exp = \cos + i\sin$ collect the amplitude of each term. For example, when $(\lambda_1, \lambda_2) = (1,2)$, we have
\begin{align}
	\exp(2i\tilde\Omega \tau_1) \exp(-2i\tilde\Omega \tau_2) = \cos{2\tilde\Omega \tau_1} \cos{2\tilde\Omega \tau_2} + \sin{2\tilde\Omega \tau_1} \sin{2\tilde\Omega \tau_2} -i \cos{2\tilde\Omega \tau_1} \sin{2\tilde\Omega \tau_2} + i\sin{2\tilde\Omega \tau_1} \cos{2\tilde\Omega \tau_2} \nonumber
\end{align}
which results in $M$ that has the form $\begin{pmatrix}
	1 & +i \\
	-i & 1
\end{pmatrix}$.
Again, we require
\begin{align}
	\abs{\sum_{\stage} \sum_{i} A_i \, f_i(b, \tau)\eval^{\tfz{\stage}}_{\tiz{\stage}} + \sum_{\substack{\stage > \stage' \\ \text{non-zero trace}}} \sum_{ij} B_{ij} \, f_i(b, \tau)\eval^{\tfz{\stage}}_{\tiz{\stage}} f_j(b', \tau)\eval^{\tfz{\stage'}}_{\tiz{\stage'}}}^2 < 1 \, .
\end{align}
Using the triangle inequality and Cauchy-Schwarz inequality, we can write
\begin{subequations}
	\begin{align}
		&\abs{\sum_{\stage} \sum_{i} A_i \, f_i(b, \tau)\eval^{\tfz{\stage}}_{\tiz{\stage}} + \sum_{\substack{\stage > \stage' \\ \text{non-zero trace}}} \sum_{ij} B_{ij} \, f_i(b, \tau)\eval^{\tfz{\stage}}_{\tiz{\stage}} f_j(b', \tau)\eval^{\tfz{\stage'}}_{\tiz{\stage'}}}^2 \nonumber \\ 
		&\leq \left(\sum_\stage \sum_i \abs{A_i}^2 + \sum_{\substack{\stage > \stage' \\ \text{non-zero trace}}} \sum_{ij} \abs{B_{ij}}^2\right) \left(\sum_\stage \sum_i \abs{f_i(b, \tau)\eval^{\tfz{\stage}}_{\tiz{\stage}}}^2 + \sum_{\substack{\stage > \stage' \\ \text{non-zero trace}}} \sum_{ij} \abs{f_i(b, \tau)\eval^{\tfz{\stage}}_{\tiz{\stage}} f_j(b', \tau)\eval^{\tfz{\stage'}}_{\tiz{\stage'}}}^2\right) \label{eq:ineq}
	\end{align}
\end{subequations}

To obtain the high-fidelity condition, we require the term on the right-hand side to be as small as possible. Since $f_i(b, \tau)$ is a periodic function composed by sine and cosine, the maximum possible value of $\abs{f_i(b, \tau)\eval^{\tfz{\stage}}_{\tiz{\stage}}} = 2$. In addition, there are 5 terms in the sum over all stages $\sum_b$, and 4 terms in the sum over different stages $\sum_{b>b^\prime}$. Also, there are 7 frequency terms in the sum over the frequencies of $A$ as given in Eq.~\eqref{eq:Ai}, and $(16+16+1)= 33$ frequency terms in the sum over the frequencies of $B$ as given in  Eq.~\eqref{eq:Bij}.  Finally, there are 4 possible terms with non-zero trace with $\stage>\stage'$: $(\stage, \stage') \in \{(5,1), (4,3), (4,2), (3,2)\}$. We can thus express the right-hand-side of the inequality~\eqref{eq:ineq} as
\begin{align}
	&  \left(\sum_\stage \sum_i \abs{f_i(b, \tau)\eval^{\tfz{\stage}}_{\tiz{\stage}}}^2 + \sum_{\stage > \stage'}\sum_\text{non-zero trace} \sum_{ij} \abs{f_i(b, \tau)\eval^{\tfz{\stage}}_{\tiz{\stage}} f_j(b', \tau)\eval^{\tfz{\stage'}}_{\tiz{\stage'}}}^2\right) \left(\sum_\stage \sum_i \abs{A_i}^2 + \sum_{\substack{\stage > \stage' \\ \text{non-zero trace}}} \sum_{ij} \abs{B_{ij}}^2\right) \nonumber \\
	&< \left(5 \cdot 7 \cdot 2^2 + 4 \cdot 33 \cdot 2^2 \cdot 2^2 \right)\left(5 \sum_i \abs{A_i}^2 + 4 \sum_{ij} \abs{B_{ij}}^2\right) = 2252 \left(5 \sum_i \abs{A_i}^2 + 4 \sum_{ij} \abs{B_{ij}}^2\right). \label{eq:AB}
\end{align}

Now, since $\tilde\omega_d = \tilde\omega_r - 2\tilde\Omega$, we can denote the ratio between the photon frequency and the driving amplitude to be $x := \frac{\tilde\omega_r}{\tilde\Omega}$. This allows us to express the frequency ratios in $(A_i)$ and $(B_{ij})$ in terms of $x$. Substituting the frequencies of Eqs.~\eqref{eq:Ai},~\eqref{eq:Bij} in terms of $x$ into the right-hand-side of Eq.~\eqref{eq:AB}, we obtain the high-fidelity condition by requiring
\begin{align}
	\frac{5}{16\left(x-2\right)^4} + \frac{9}{\left(x-2\right)^4} + \frac{5}{16\left((x-2)^2 - 1\right)^2} + \frac{21 + 42(x-2)^2 + 21(x-2)^4}{\left((x-2)^2 - 1\right)^4} < \frac{1}{2252}.
\end{align}
From the above, we obtain $x > \varrho \approx 18.269$, which gives us the boundary of the high-fidelity region of the spin qubit:
\begin{align}
	2 \tilde\Omega  < \frac{2 \tilde\omega_r}{\varrho}.
\end{align}
\end{widetext}

\newpage

\bibliographystyle{apsrev4-2}
\bibliography{GZ_Yang_References}

\end{document}